\documentclass[useAMS,fleqn,usenatbib,usegraphicx]{mn2e}
\usepackage{amsmath}

\def\teff{${\rm T}_{\rm eff}$}

\def\av{$A_{\rm V}$}

\title[Finding rare objects]{Finding rare objects and building
  pure samples: Probabilistic quasar classification from low resolution Gaia
  spectra}

\author[C.A.L.\ Bailer-Jones et al.]
{C.A.L.\ Bailer-Jones$^1$\thanks{Email: calj@mpia.de}, 
K.W.\ Smith$^1$, C.\ Tiede$^1$, R.\ Sordo$^2$, A.\ Vallenari$^2$\\
$^1$Max-Planck-Institut f\"ur Astronomie, K\"onigstuhl 17, 69117
Heidelberg, Germany\\
$^2$INAF, Osservatorio di Padova, Vicolo Osservatorio 5, 35122 Padova, Italy\\
}

\begin{document}

\date{submitted 7 August 2008; resubmitted 18 September 2008}

\maketitle

\label{firstpage}

\begin{abstract}
  We develop and demonstrate a probabilistic method for classifying
  rare objects in surveys with the particular goal of building very
  pure samples. It works by modifying the output probabilities from a
  classifier so as to accommodate our expectation (priors) concerning
  the relative frequencies of different classes of objects.  We
  demonstrate our method using the {\em Discrete Source Classifier}, a
  supervised classifier currently based on Support Vector Machines,
  which we are developing in preparation for the Gaia data
  analysis. DSC classifies objects using their very low resolution
  optical spectra.  We look in detail at the problem of quasar
  classification, because identification of a pure quasar sample is
  necessary to define the Gaia astrometric reference frame. By varying
  a posterior probability threshold in DSC we can trade off sample
  completeness and contamination. We show, using our simulated data,
  that it is possible to achieve a pure sample of quasars (upper limit
  on contamination of 1 in 40\,000) with a completeness of 65\% at
  magnitudes of G=18.5, and 50\% at G=20.0, even when quasars have a
  frequency of only 1 in every 2000 objects.  The star sample
  completeness is simultaneously 99\% with a contamination of 0.7\%.
  Including parallax and proper motion in the classifier barely
  changes the results. We further show that not accounting for class
  priors in the target population leads to serious misclassifications
  and poor predictions for sample completeness and contamination. We
  discuss how a classification model prior may, or may not, be
  influenced by the class distribution in the training data. Our
  method controls this prior and so allows a single model to be
  applied to any target population without having to tune the training data and
  retrain the model.
\end{abstract}


\section{Introduction \label{introduction}}

Finding rare objects is hard, for two reasons. First, we expect to
have to look at many objects before we encounter one, and second, we
may not even recognise it even when we do. The reason for this is that
the prior probability that any one object is of the rare class,
$P(C_{\rm rare})$, is very small. So even if it has very
characteristic features, i.e.\ the likelihood of the data given the
rare object, $P({\rm Data} \,|\, C_{\rm rare})$, is high, the
posterior probability, $P(C_{\rm rare} \,|\, {\rm Data}$), could still
be low.

A survey for rare objects obviously requires a very discriminating
classifier, but we could assist it by modifying the class
probabilities. If we raised the prior, $P(C_{\rm rare})$,
we are more likely to find the rare objects, but it is inevitable
that we will then incorrectly classify other objects as being of the
rare class.  Depending on our goals, there may be a
satisfactory balance between maximizing the expected number of true
positives (sample completeness) and minimizing the expected number of
false positives (sample contamination).  Here we present a method for
achieving an optimal balance and a correct prediction of this balance
which, although simple, is not trivial and has important implications.

We illustrate our method in the context of the problem of detecting
quasars in the Gaia survey based on their low resolution ($R \simeq
30$) optical spectra.  Gaia is an all-sky astrometric and
spectrophotometric survey complete to $G=20$, expecting to observe
some $10^9$ stars, a few million galaxies and half a million
quasars. Its primary mission is to study Galactic structure by
measuring the 3D spatial distribution and 2D kinematic distribution of
stars throughout the Galaxy and correlating these with stellar
properties (abundances, ages etc.) derived from the spectra. With
astrometric accuracies as good as 10\,$\mu$as, Gaia cannot be
externally calibrated with an existing catalogue. Instead it must
observe a large number of quasars over the whole sky with which to
define its own reference frame.  This quasar sample must be very clean
(low contamination).  (The quasar sample is also, of course, of intrinsic
interest.)

We present our Gaia classification model, the {\em Discrete Source
  Classifier} (DSC), and report its classification performance using
simulated data.  We will show how, using our probabilty modification
approach, we can use this to build pure quasar samples, at the (acceptable)
loss of sample completeness.

{\bf Related work}. One of the most comprehensive search for quasars
to date is that done with the SDSS. Richards et al.~\cite{richards02}
defined a colour locus in the {\it ugriz} space to identify objects
for spectroscopic follow up and estimated the contamination rate of
their photometric selection to be 34\%.  Using the spectroscopy of all
point sources taken in SDSS stripe 82, Vanden Berk et
al.~\cite{vandenberk05} assessed the completeness of the Richards et
al.\ selection at 95.7\% for sources brighter than $i=19.1$. Later,
Richards et al.~\cite{richards04} trained a photometric classifier
(based on kernel density estimation) on a set of 22,000
spectroscopically identified quasars and used this to build a sample
of 100\,000 quasars brighter than $g=21.0$.  For the unresolved UV
excess quasars in this sample, they estimated the completeness to be
94.7\% down to $g$=19.5, with a contamination of just 5\% down to
$g$=21.0. Suchkov et al.~\cite{suchkov05} and Ball et
al.~\cite{ball06} both use decision trees to classify objects in the
SDSS photometric catalogue, by training on objects with spectroscopic
classifications from SDSS. Gao et al.~\cite{gao08} used support vector
machines and Kd-tree nearest neighbours to classify
spectroscopically-confirmed SDSS and 2MASS objects as quasars or stars
using just the photometry, with stars and quasars present in
equal proportions. With both methods they obtained
contamination rates of a few percent (averaged across both classes).


{\bf Outline}. We start in section~\ref{theory} by presenting our general prior
modification method. In section~\ref{model} we present the Gaia-DSC
classifier and the data used to train and test it. The performance of this and
the application of our method are given in section~\ref{results} and
discussed in section~\ref{discussion}, where we also discuss the
interpretation of the probabilities and some of the limitations of
this work. We conclude in section~\ref{conclusions}.

{\bf Definitions and notation}.  We use the term {\em class fractions}
to mean the relative numbers of objects of each class in a data
set. The {\em nominal model} refers to the classifier ``as is'', i.e.\
the output probabilities without any modifications. In
contrast, in the {\em modified model} we have modified these outputs
to be appropriate to a situation in which there would be very
different class fractions, e.g.\ quasars very rare
(section~\ref{probmod}).  The subscript $i$ refers to true classes and
the subscript $j$ to model output classes.  $f^{train}_i$,
$f^{test}_i$ and $f^{mod}_i$ refer to the class fractions of class
$i$ for the training data, test data and effective test data
respectively. 
They are normalized, $\sum_i f^{train}_i = \sum_i f^{test}_i = \sum_i f^{mod}_i = 1$.
This {\em effective test data set} is one which we
don't actually have, but reflects a target population with different
class fractions, e.g.\ quasars very rare. We make predictions of
the performance on this by using the actual
test data but by modifying the performance equations
(section~\ref{modcalcs}).  In the figures especially, lower case class
names, e.g.\ {\tt quasar}, denote estimated classes and upper case
names, e.g.\ {\tt STAR}, true classes. Thus $P({\tt quasar} | {\tt
  STAR})$ means ``the DSC-assigned quasar probability given that the
source is truly a STAR''. This refers to DSC outputs for objects
with known classes (e.g.\ in a test set). This is not to be confused
with the notation for the DSC outputs in the general case, $P(C_j |
x_n, \theta)$, which means ``the probability which DSC model $\theta$
assigns to class $C_j$ for object $x_n$''. ``C\&C'' stands for
``completeness and contamination''.

\section{Using and modifying probabilities for classification}\label{theory}

\subsection{The issue of priors}

The probabilities and assigned classes from any classifier depend not
only on the evidence in the data, but also on the prior
probabilities. In this context, the prior is the class probabilities
before you look at the data.  Priors are always present (whether you
accept the Bayesian philosophy or not), but they are not explicit in
many models, so we may not know what they are.  This is a problem,
because if the model priors are inappropriate (e.g.\ equal probability
of star and quasar) this will translate into poor posterior
probabilities and thus class assignments.  The method we present here
allows us to replace priors (even if implicit) with something more
appropriate, e.g.\ quasars much rarer than stars.

\subsection{Building samples and assessing completeness and
  contamination\label{sample}}

DSC assigns to every data vector (e.g.\ spectrum), $x_n$, a
probability, $P({\tt quasar})$, that it is a quasar, and likewise for
the other classes.
The classes are exclusive and exhaustive, so the probabilities sum to
unity.  The classifier is not perfect, so generally $P({\tt quasar} |
{\tt QUASAR}) < 1$ and $P({\tt quasar} | not\,{\tt QUASAR}) > 0$.

We could simply assign an object to that class for which the DSC
output is largest (the most probable class). However, if the different output
probabilities for an object were all similar this would not be a confident
classification.  A more secure basis for building pure samples is to
select objects only if their probability is above some threshold: the
higher this is, the lower the sample contamination, but the smaller
also its completeness. 

We assess the performance using a labelled test set. To build a {\em sample}
for class ${\tt class}$, we select objects which have a DSC output
$P({\tt class}) > P_{\rm t}$, where $P_{\rm t}$ is a user-defined
threshold (and may be different for each class).  The sample {\bf
  completeness} is the number of objects in the sample which are
truly of that class divided by the total number of objects of that
class in the test set.  The {\bf contamination} is the number of
objects in the sample which are truly of other classes divided by the
total number of objects in the sample. For ${\tt class} = j$
\begin{eqnarray}
{\mathrm completeness}_j &=& \frac{ n_{i=j,j} }{ N_i } \nonumber \\
{\mathrm contamination}_j &=& \frac{ \sum_{i \neq j} n_{i,j} }{  \sum_i n_{i,j}} 
\label{eqn:compcont}
\end{eqnarray}
where $n_{i,j}$ is the number of objects of (true) class $i$ in the
sample selected for (output) class $j$, and $N_i$ is the total number
of objects of (true) class $i$ in the test set. (There are
various other diagnostics one could use, such as the ROC curve or precision rate.)
These equations give us predictions of the completeness and
contamination for a new (unlabelled) data set, insofar as we believe
it to have the same class fractions as the training data (that's our
prior).

\subsection{Model-based class priors}\label{mbp}

All classifiers include a prior, whether explicit or not. We need to
know this prior for two reasons.  First, we would like to know what
assumption our model is {\em actually} making (and not what we {\em
  suspect} it is making). Second, we would like to change this prior
to something which is appropriate to the problem at hand. Here we
discuss what the prior is, how the training data may influence it, and
how to calculate it post hoc from a trained model.

\subsubsection{Bayes and priors}

The outputs from a trained classifier when presented with data $x_n$
are $P (C_j | x_n, \theta)$, the probability that
the data is of class $C_j$ given the data and the model,
$\theta$. This latter quantity reflects both the architecture of the
model and the training data set used to fix its internal parameters. 
We can think of this output as a posterior probability and write it using
Bayes' theorem
\begin{equation}
P(C_j | x_n, \theta) = \frac{P(x_n | C_j, \theta) P(C_j | \theta)}{P(x_n | \theta)}
\label{bayes}
\end{equation}
The term $P(x_n | C_j, \theta)$ is the {\em likelihood} of the data
given the class and model. The term $P(C_j | \theta)$ is the {\em
  prior probability} that, given our model, an object is of class
$C_j$. Bayesian statistics deals with updating probabilities based on
new data: the prior reflects our knowledge (based on some other data)
before we look at the new data.  In the present context, the prior
$P(C_k={\tt quasar} | \theta)$ is the probability that any one object
in our survey is a quasar, before we look at its spectrum.  We always
have some prior information, e.g.\ with Gaia the fact that it is an
all sky survey to G=20.0. If we know them, we could even treat the
magnitude and Galactic latitude as prior information.

\subsubsection{What are the classifier priors and how does the training data influence them?}\label{traininf}

Given that we can make the decomposition in equation~\ref{bayes}, it
follows that all classification models must possess a prior on the
class probabilities. In some models, for example linear discriminant
analysis or Gaussian mixture models, this prior is explict and so can
be controlled. But in many others, such as neural networks or support
vector machines, it is not explicit. (See a standard text on machine
learning for details of these methods, e.g.\ Hastie et
al.~\cite{hastie01}.) In particular, it may depend on the class
fractions in the training data.

Take, for example, a standard neural network regression model which is
trained by minimizing an error function over the whole data set.  If
we trained this on 1000 stars and just one quasar, it will learn
to recognise stars much better than quasars, because in minimizing
the error it hardly has to worry about fitting the lone quasar. If we
changed the training data (class fractions), 
the model and thus the classifications would change. Other
regression models are influenced by the class fractions in
different ways, or not at all.  Given this dependence on the model and
data, we refer to the priors as {\em model-based priors}, and the
notation $P(C_j | \theta)$ reminds us of this.

This issue of class fractions influencing the model performance is
well-known in the machine learning literature, where it is referred to
as the problem of ``class imbalance'' or ``imbalanced data sets''.  It
has been demonstrated to influence neural networks, support vector
machines and classification trees (e.g.\ Shin \& Cho 2003, Visa \&
Ralescu 2005, Weiss 2004). But how, exactly, do the class fractions
affect the classifications and, more specifically, the model-based
priors? We might think that in the above example the ratio of the star
to quasar prior probabilities implicit in the model is 1000 to 1, but
this is generally not the case\footnote{If this were the case, then we
  might be tempted to address the class imbalance problem by changing
  the training data to have class fractions equal to our priors. But
  if we had just 10\,000 training vectors, we could then have only 10
  quasars, making it hard for the classifier to correctly classify
  quasars. We demonstrate this later (section~\ref{reduceqsos}).},
because it depends on the model and how it is trained. The bottom line
is that, in general, the model-based prior is not equal to the class
fractions in the training data.

\subsubsection{Calculating the model-based priors\label{mbpnom}}

We can calculate the model-based priors, $P(C_j | \theta)$, directly
from the trained model via the marginalization equation
\begin{equation}
P(C_j | \theta) = \sum_{n=1}^{n=N_{test}} P(C_j | x_n, \theta) P(x_n | \theta)
\label{margin}
\end{equation}
where the sum is taken over all $N_{test}$ objects in the test data
set.  The first term in the sum is the posterior probability.  The
second term is the probability that we draw object $x_n$ from the test
data set, which is $1/N_{test}$.  Hence the prior is simply the
average of the posterior probabilities. It might seem strange that the
prior can be calculated from the posterior. Yet because the sum is over all
objects in the test set, regardless of their true class, we can
think of this summation as eliminating information on individual
objects, leaving us with what the model probability is for class $C_j$
in the absence of specific data. This is the prior.  If we had
three classes equally represented in the data, and the classifier were
perfect, the prior for each class would be $1/3$. Different class
fractions and non-perfect classifiers will give different
results.\footnote{Note that the prior calculation assumes the test 
  set has the same class fractions as the training set. Interestingly,
  because the true classes don't appear in the equation, we don't
  actually need labels on the individual objects in order to
  calculate the model-based prior.}

In section~\ref{results} we will compare the
model-based priors with the training data class fractions.

\subsection{Modifying the posterior probabilities to account for modified
  class fractions or priors\label{probmod}}

Typically we train a classification algorithm on more or less equal
class fractions, because this helps the algorithm to reliably identify
the class boundaries. This is our nominal model, $\theta^{nom}$.  Yet
the target population to which we want to apply our classifier may
have different class fractions (e.g.\ quasars very rare), in
which case the classifier may make poor predictions.  The underlying
issue here is that the model-based priors of our nominal model differ
from our target priors.

To correct for this we would ideally just replace the original
(nominal) prior with our modified prior. But we cannot do this
directly because the model priors are not explicit. Instead we modify
the output (posterior) probabilities to give us the ``modified
model'', $\theta^{mod}$.  From inspection of equation~\ref{bayes} we
see we may achieve this by dividing the posterior by the nominal prior
and then multiplying it by the modified prior.  The nominal priors can
be calculated using equation~\ref{margin}. But we cannot use this
equation to calculate the modified priors because we don't yet have
the posteriors from the modified model. We therefore approximate the
modified priors using the class fractions. Let $P^{nom}(C_j | x,
\theta^{nom})$ be the output (posterior) probability from the nominal
model, i.e.\ that trained on the data set with class fractions
${f^{train}_i}$. The modified probability appropriate for a target
population with class fractions $f^{mod}_i$ we define as
\begin{equation}
P^{mod}(C_j | x_n, \theta^{mod}) = a_n \, P^{nom}(C_j | x_n, \theta^{nom}) \, \frac{f^{mod}_{i=j}}{f^{train}_{i=j}}
\label{eqn:modprob}
\end{equation}
where $a_n$ is a normalization factor which ensures that $\sum_j
P^{mod}(C_j | x_n, \theta^{mod}) = 1$ for each object $n$.  The factor
${f^{mod}_{i=j}}/{f^{train}_{i=j}}$ is equivalent to changing the
prior in equation~\ref{bayes} and has the expected impact on the
posterior.  

Following the discussion in section~\ref{mbp} we do not necessarily
expect the class fractions to be a good proxy for the priors, but for
now it's the best we can do. (We will later check how good the
approximation is.)

We illustrate the above with a simple example, a two-class model
trained on equal numbers of stars and quasars, i.e.\ $f^{train} = (0.5, 0.5)$. Let's suppose an object is classified with the
nominal model to give $P^{nom}(C_{star} | x_n, \theta^{nom}) = 0.2$
and therefore $P^{nom}(C_{quasar} | x_n, \theta^{nom}) = 0.8$. If we
want $f^{mod} = (0.9,0.1)$ (stars nine times more common than
quasars), then from equation~\ref{eqn:modprob} we get
\begin{flalign*}
P^{mod}(C_{star} | x_n, \theta^{mod}) = a_n \, 0.2 \times 0.9/0.5 = a_n \, 0.36 = 0.69 \\
P^{mod}(C_{quasar} | x_n, \theta^{mod}) = a_n \, 0.8 \times 0.1/0.5 = a_n \, 0.16 = 0.31
\end{flalign*}
the final values being those after normalization.  We see how the
prior can have a significant effect on our inference.

It is tempting to think of these modified probabilites as those we
would have obtained {\em if} we had trained the model on the modified
class fractions. But this is generally {\em not} true, because what we
are doing in equation~\ref{eqn:modprob} is changing the priors (via
the proxy of the class fractions) and this is not equivalent to
changing the training data (as we will demonstrate later).

\subsection{Predicting performance and model-based priors for an
  effective test set using the nominal test set }\label{modcalcs}

\subsubsection{The completeness and contamination calculation}\label{modcc}

The equations for the completeness and contamination
(eqn.~\ref{eqn:compcont}) depend on the actual
number of objects correctly and incorrectly classified and were
calculated from the test data set, which by definition has class
fractions $f^{test}_i$. However, we expect to apply the modified model
to a target population with different class fractions (this was the
whole point of modifying it). To estimate the C\&C, it is not necessary
to create a new test data set. Instead, we can modify the
equations to reflect the target class fractions, $f^{mod}$, namely
\begin{eqnarray}
{\mathrm completeness}_j &=& \frac{ \left(\frac{f^{mod}_{i=j}}{f^{test}_{i=j}}\right) n_{i=j,j} }{ \left(\frac{f^{mod}_{i=j}}{f^{test}_{i=j}}\right) N_i } = \frac{ n_{i=j,j} }{ N_i } \nonumber \\
{\mathrm contamination}_j &=& \frac{ \sum_{i \neq j} \left(\frac{f^{mod}_i}{f^{test}_i}\right) n_{i,j} }{  \sum_i \left(\frac{f^{mod}_i}{f^{test}_i}\right) n_{i,j}}   
\label{eqn:mod_compcont}
\end{eqnarray}
This effectively changes the number of objects in each true class.
The quantities $n_{i,j}$ and $N_i$ refer to the number of objects in
samples obtained from the modified model applied to the actual test
data set.  Note that the {\em equation} for the completeness remains
unchanged, yet the {\em value} of the completeness will generally
change because the posterior probabilities have changed and so
$n_{i=j,j}$ changes. We emphasise that the above equation is
independent of the procedure to modify the probabilities described in
section~\ref{probmod}. It is only to be used to make predictions for
an {\em effective} test set. To calculate the completeness and
contamination for an {\em actual} test set which already has the
modified class fractions, we use equation~\ref{eqn:compcont}.

\subsubsection{The model-based prior calculation}\label{modmbps}

Once we have the posteriors from the modified model we can calculate
its model-based priors from the test set using
equation~\ref{margin}. But just as for the C\&C calculation, we must
change the effective number of objects in each true class to represent
the modified class fractions. Thus in equation~\ref{margin} we should
use
\begin{equation}
P(x_n | \theta^{mod}) = \left(\frac{f^{mod}_i}{f^{test}_i}\right) \frac{1}{N_{test}}
\label{eqn:px.thetamod}
\end{equation}
where $x_n$, being an object in the (labelled) test set, has known
true class, $i$. If a class is ten times rarer in the modified
population, then $P(x_n | \theta^{mod})$ is reduced by a factor of ten compared to before.

In principle we could now use these model-based priors to recalculate
the posteriors for the modified model (without using the approximation
in equation~\ref{eqn:modprob}), but we don't do this in the present
article. Rather we just report them as a check on what the prior
really is.

\section{The Discrete Source Classifier}\label{model}

The Discrete Source Classifier (DSC) is the algorithm we are
developing as part of our work in the Gaia Data Processing and
Analysis Consortium (DPAC) (O'Mullane et al. \cite{omullane07}).  DSC
will classify objects based primarily on low resolution prism spectra
(so-called ``BP/RP'' spectra), but supplemented with parallaxes and
proper motions and photometric variability where available.  Here we
restrict ourselves to the three classes ``star'', ``galaxy'' and
``quasar''.  After DSC in the pipeline come several algorithms for
estimating astrophysical parameters, e.g.\ stellar parameters. For
more details of the classification system, see Bailer-Jones
\cite{cbj05}.

Gaia detects objects in a very broad band, G, which has roughly the
same wavelength range as the BP/RP instrument. In common with the
Gaia detection strategy, DSC only operates on point sources (i.e.\
makes no use of morphological information).  In this paper we present
results for simulations at both G=20.0 and G=18.5.

Here we describe the underlying classifier, the source libraries and
the Gaia simulated data used to train and test DSC.

\subsection{The SVM classification algorithm}

DSC is presently based on a Support Vector Machine (SVM)
(e.g.\ Cortes \& Vapnik~\cite{cortes95}, Burges~\cite{burges98}),
using the libSVM implmentation (Chang \& Lin~\cite{libsvm}).  An SVM
is a supervised learning algorithm which finds the boundary which
``optimally'' separates two classes of objects.  In contrast to many
other classifiers, SVMs are defined only by those training objects
which lie close to the boundary, the so-called {\em support vectors}.
The basic algorithm is linear, but by using the ``kernel trick'' to
implicitly map data into a higher dimensional space, it achieves a
nonlinear mapping between the input data (the spectrum) and the output
classes.  The SVM is trained by minimizing the number of
misclassifications and using a regularizer for complexity control.

SVMs are fundamentally non-probabilistic: they simply define a
boundary and assign a class ($C_1$ or $C_2$) depending on which side
of the boundary an object falls.  Probabilities may be derived from
the distance, $f$, of an object from the boundary, in the present case
by fitting a sigmoidal function to $P(C_j=C_1 | f)$
(Platt~\cite{platt99}).

The basic SVM is a two-class classifier.  libSVM actually trains
$K(K-1)/2$ one-against-one classifiers (for a $K$ class problem) and
uses the pairwise coupling method described in Wu et al.~\cite{wu04}
(their algorithm 2) to produce normalized probabilities for $K$
classes.

One of the advantages of SVMs over many other nonlinear classifiers
(e.g.\ neural networks) is that the SVM objective function is convex,
so the training has a unique solution. However, the model has two hyperparameters,
the regularization parameter (the ``cost'') and the scale length in
the radial basis function (RBF) kernel we use.  We optimize these (``tune'')
using a Nelder-Mead algorithm. As this involves retraining the SVM for
each combination of hyperparameters, it is a time-consuming process.

The main concepts presented in this paper are not dependent on the
choice of classifier. SVM is not perfect, not least because its
underlying philosophy of not modelling the data distribution means
that probabilties do not arise naturally. It is, nonetheless, a
competitive model in terms of classification error, and in tests we
have done it has performed as well as or better than several
alternatives, including multilayer perceptrons, RBF networks, boosted
trees and mixture models (Elting \& Bailer-Jones~\cite{elting07}).

\subsection{Source spectral libraries}

\begin{figure}
\begin{center}
\includegraphics[width=0.30\textwidth]{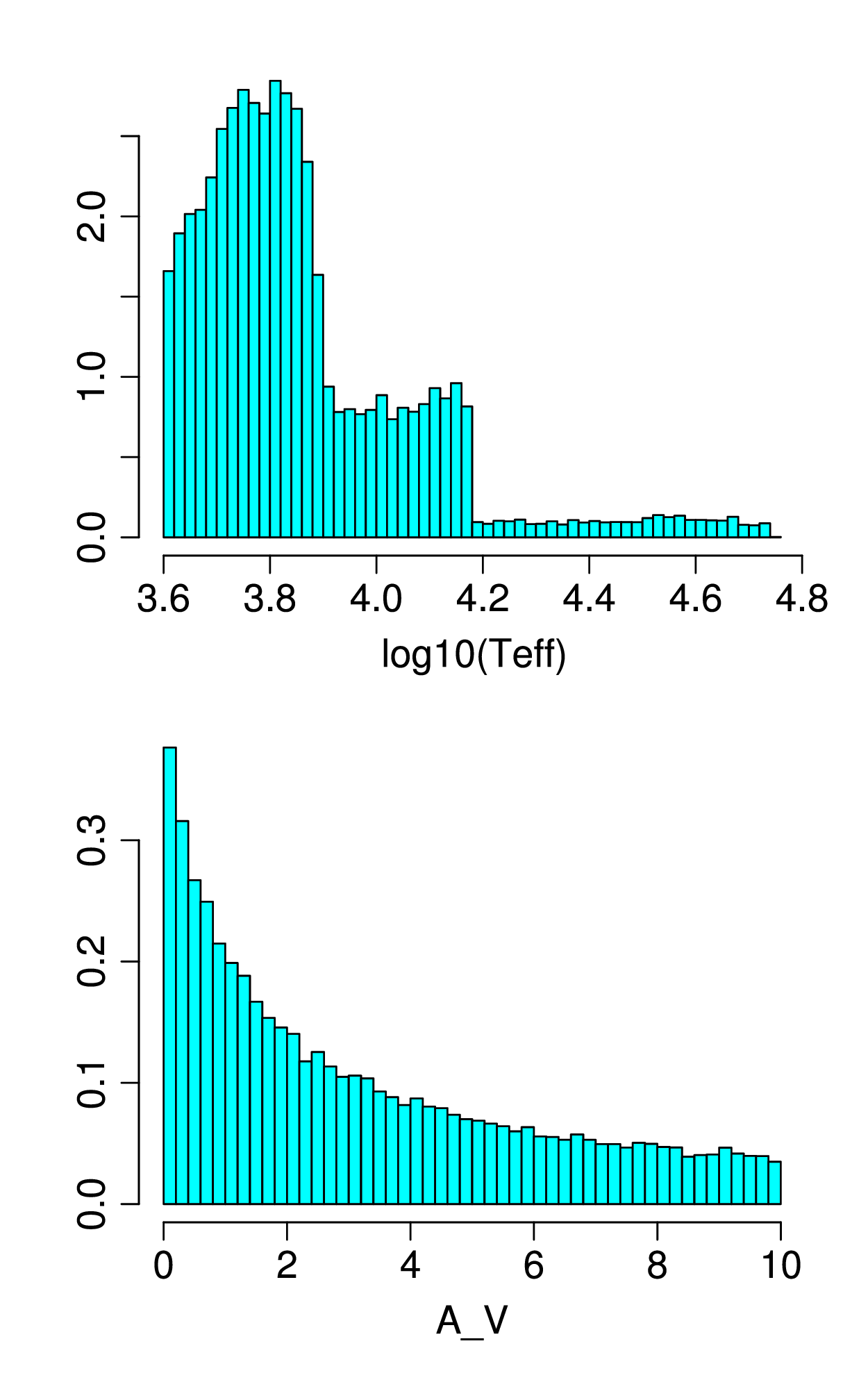}
\caption[Stellar AP distribution]{The distribution of stellar
  parameters in the training and testing data sets\label{large_star_aps}}
\end{center}
\end{figure}

Libraries of spectra are being assembled or created for Gaia
purposes. Our stellar libraries are based on the Basel (Lejeune et
al.~\cite{lejeune97}) and MARCS (e.g.\ Gustafsson et
al.~\cite{gustafsson08}) libraries plus an OB star library from J.-C.\
Bouret. Together these span the full range of effective temperature,
metallicity and surface gravity.  The galaxy library is described by
Tsalmantza et al.~\cite{tsalmantza08} (see also Tsalamantza et
al.~\cite{tsalmantza07}). It is based on synthetic spectra derived
from galaxy evolution models with several astrophysical parameters, in
particular those which describe the star formation history and the
redshift.  The quasar library is also synthetic, with each spectrum
uniquely defined by the three parameters continuum slope, $\alpha$,
emission line equivalent width, EW, and redshift, $z$ (Claeskens et
al.~\cite{claeskens06}). Their values are chosen at random from a
distribution which is flat in redshift over the range 0--5.5, flat in
$\alpha$ from $-4$ to $+3$ and exponentially decreasing in EW from 0
to 100\,000\,\AA.  We further simulated the galaxy and stellar spectra
at random values of interstellar extinction over the range $A_V
=$\,0--10 (with fixed extinction law $R_V = 3.1$). The main stellar
parameters which influence the potential contamination with the
quasars are \teff\ and \av; their
distribution is plotted in Fig.~\ref{large_star_aps}. The
distributions may be a little artificial (the three ``steps'' reflect
the three libraries), but they are sufficient for the purpose of this
article.

This data set is obviously rather limited (e.g.\ purely synthetic
data, no extinction in the quasars, no emission line stars, no strong
galaxy emission). Our main goal at this stage is not to produce a data
set tuned to the real Gaia sky. We discuss the issue of training sets
more in section~\ref{discussion}.

\subsection{Gaia simulated spectra}

To build and test the processing pipeline, the DPAC has built a
sophisticated data simulator (Luri et al.~\cite{luri05}).  For each
object in the spectral libraries we simulate its BP/RP spectrum,
parallax and proper motion, including all sources of random
error. Gaia observes every part of the sky between 40 and 200
times over five years (depending on ecliptic latitude).  The average
number of spectra for an object in an end-of-mission stack is 72,
which we use here.  Our simulated data are the so-called cycle 3 data
(Sordo \& Vallenari~\cite{sordo08}).

\begin{figure}
\begin{center}
\includegraphics[width=0.48\textwidth]{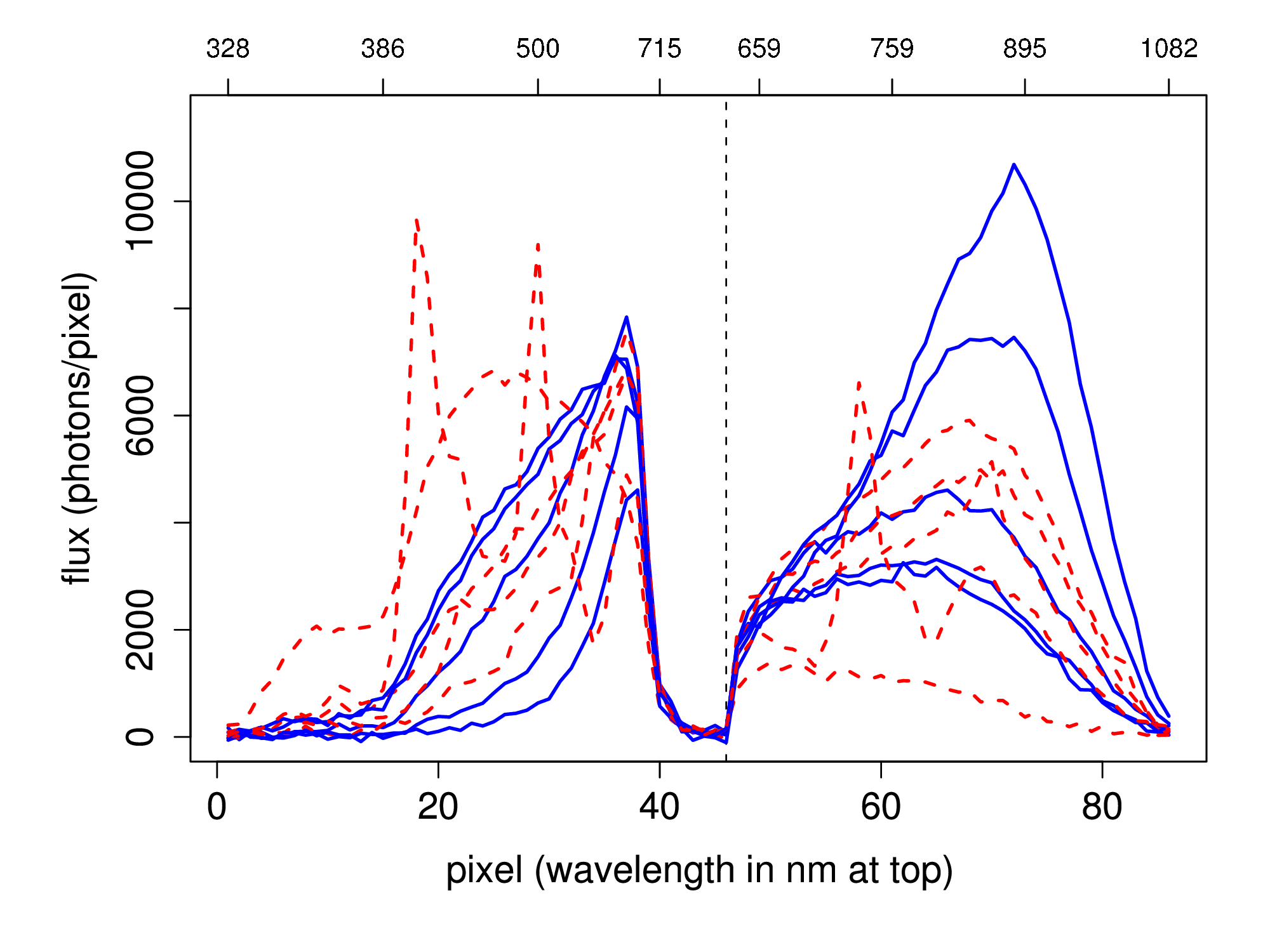}
\caption[Example stellar and quasar spectra]{Random selection of five
  stars (blue solid lines) and five quasars (red dashed lines)
  (G=18.5, noise included). The first 46 pixels are the blue channel (Blue
  Prism, BP), the remaining 40 the red channel (Red Prism, RP),
  separated by a dashed line.\label{fig:random_stars_quasars}}
\end{center}
\end{figure}

\begin{figure}
\begin{center}
\includegraphics[width=0.45\textwidth]{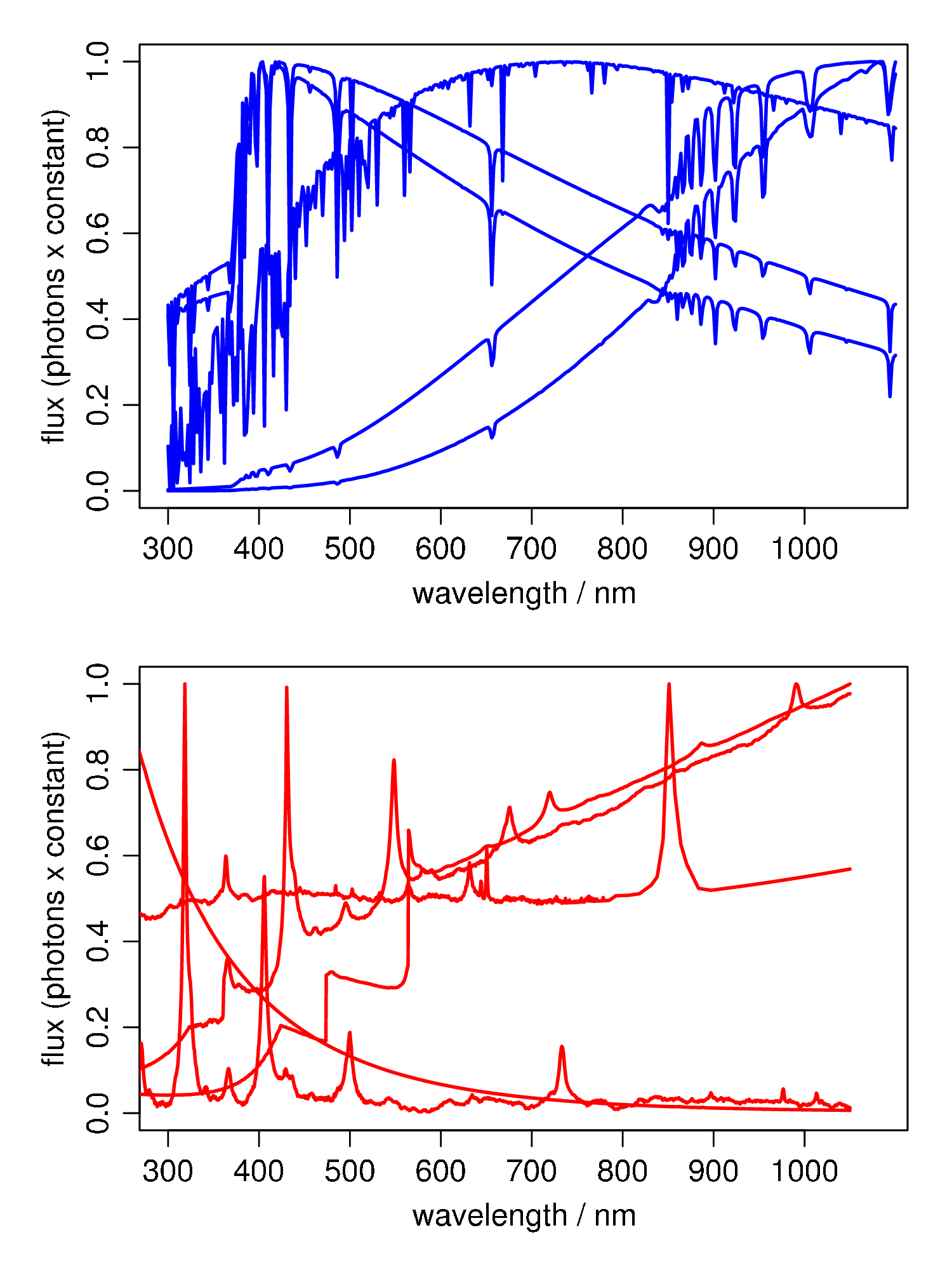}
\caption[Example stellar and quasar spectra]{The (noise-free) library spectra for the
  stars (top panel) and quasars (bottom panel) shown in
  Fig.~\ref{fig:random_stars_quasars}. The photon flux has been
  normalized in each case to have a maximum of
  1.0.\label{fig:random_stars_quasars_inputspectra}}
\end{center}
\end{figure}

Gaia prism spectra are obtained in two channels, one for the blue
called BP and one for the red called RP, with cut-offs defined by the
(silver) mirror response in the blue, CCD QE in the red, and bandpass
filters in the middle. After removal of low sensitivity regions, BP
ranges from 338 to 998\,nm over 46 pixels and RP from 646 to 1082\,nm
over 40 pixels. We use all 86 pixels as separate inputs to the
classifier (we do not attempt to combine the wavelength overlapped
region).
The dispersion varies strongly with wavelength from 3 nm/pixel (blue
end) to 30 nm/pixel (red end).  The instrumental point spread function
(PSF) is significantly broader than the pixel size, and in one conservative
estimate there are are only 18 independent measures in BP/RP
(Brown~\cite{brown06}). Example spectra are shown in
Fig.~\ref{fig:random_stars_quasars}. The dominant noise source is
Poisson noise from the source, so the SNR is approximately the square
root of the flux. (At G=20, the fluxes are four times smaller and the
SNR is about half.)  The corresponding library spectra (i.e.\ prior to
being put through the Gaia instrument simulator) are shown in
Fig.~\ref{fig:random_stars_quasars_inputspectra}. The original stellar
library spectra are calculated at a resolution of 0.1\,nm and show
many absorption lines, so to avoid confusion, we just plot every
20$^{th}$ pixel. Note how much information is lost -- especially
concerning the spectral lines -- in BP/RP.

\subsection{Simulated astrometry}

\begin{figure}
\begin{center}
\includegraphics[width=0.25\textwidth]{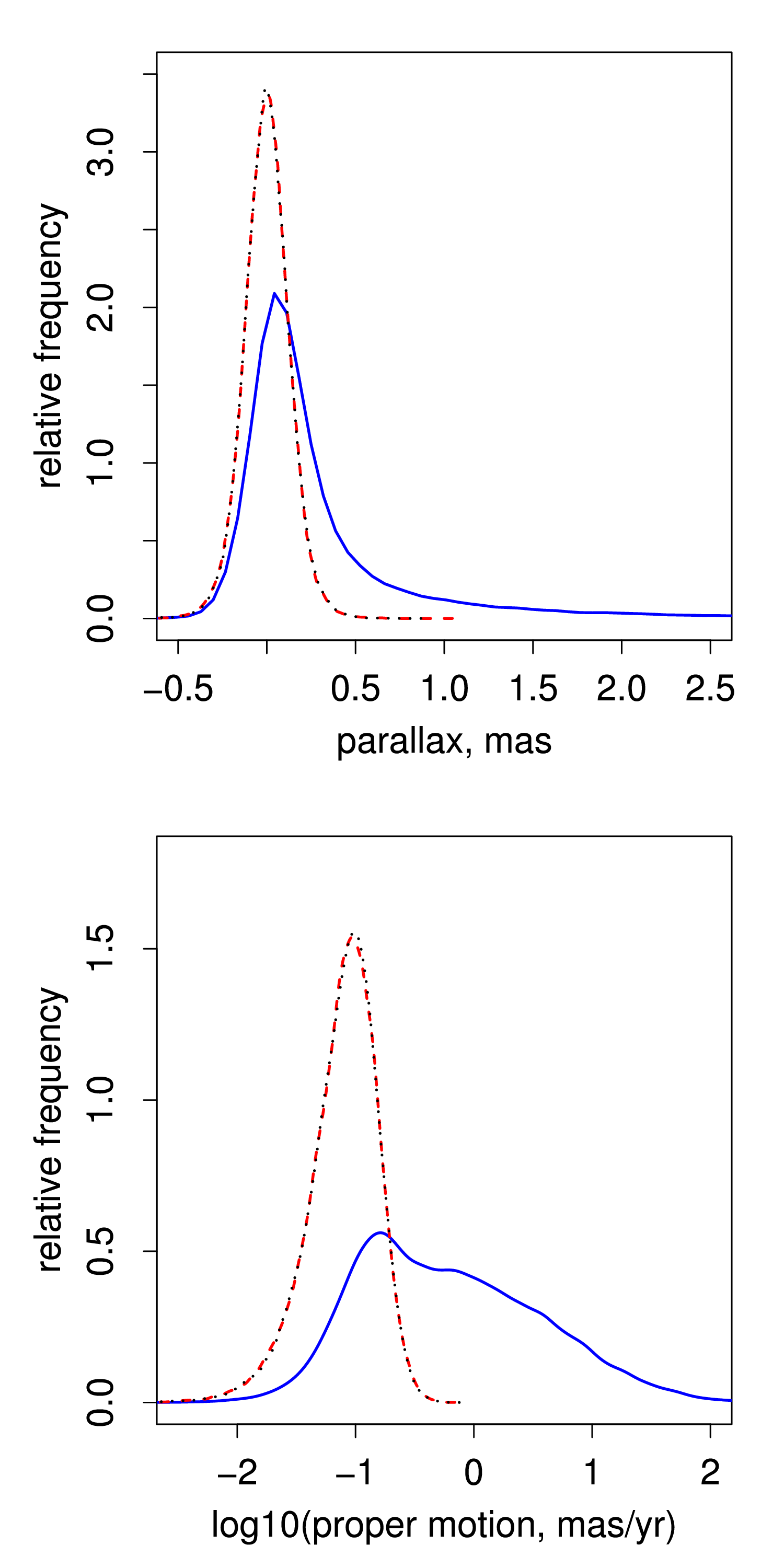}
\caption[Proper motion and parallax density]{Distribution of the simulated astrometry.
Blue/solid=star, red/dashed=quasar, black/dotted=galaxy (which is indistinguishable from the quasars).
\label{fig:g18.5_astrometry_density}}
\end{center}
\end{figure}

We included the astrometry in our classifiers in the expectation that
it could help distingiush galactic and extragalactic objects.
Fig.~\ref{fig:g18.5_astrometry_density} shows the distribution of the
simulated astrometry (with noise) 
for G=18.5. Astrometry for the extragalactic objects is just zero plus
noise. Note that this causes negative parallaxes (which are preserved
in the Gaia astrometric data processing for model fit assessment
purposes).  Stellar parallaxes, $\varpi$, are simulated in a way to
ensure consistency with the apparent and absolute magniutudes
and stellar parameters.  The stellar proper motions are drawn
at random from a zero mean Gaussian with standard deviation $10
\varpi$/yr, which simiulates a distance-independent space velocity
distribution with a standard deviation of 50\,km/s.

At G=18.5, the Gaia parallax error is typically 0.1\,mas and the
proper motion error around 0.14\,mas/yr (both dominated by source
photon statistics). The errors are twice as large at G=20.0.

\subsection{Train and test data}

The input data for our classifiers is BP/RP plus astrometry (although
it turns out that inclusion of astrometry hardly improves the
results).  Models are trained and tested on noisy data and a single
G-band magnitude is used in each experiment. 

Train and test data are drawn at random (without replacement) from a
common data set.  Unless stated otherwise, the training data set
comprises 5000 objects from each of the three classes and the test
data 60\,000 objects of each class. The reason for the large number in
the test set will become clear in section~\ref{g185nolowEWQSO}.

\section{Results}\label{results}

\subsection{Overview of the experiments and explanation of the figures}

We present results for three experiments
\begin{enumerate}
\item[(1)] objects at G=18.5,
\item[(2)] as (1), but in which quasars with emission line equivalent widths less
  than 5000\,\AA\ have been removed from the training data set, leaving 2901
  quasars (the test set is unchanged),
\item[(3)] as (2), but for G=20.0.
\end{enumerate}

We show results for both the ``nominal'' and ``modified'' cases.  In
the nominal cases, the class fractions are those given by the data
sets themselves, namely $f^{train} = f^{test} = (1,1,1)$ for (star,
quasar, galaxy) respectively, except for the training data in cases 2
and 3 which has $f^{train} = (1,0.58,1)$.  (We list here the class
fractions prior to normalization.) In the modified cases, the DSC
output probabilities are modified (as described in
section~\ref{probmod}) to accommodate different expected class
fractions (priors). Quasars are made 1000 times rarer, so $f^{mod} =
(1,1/1000,1)$. This is the order of the star/quasar ratio we expect
for Gaia at G=20. (In comparison, a larger ratio is found in SDSS --
as high as 0.015 at $g=18.5$ depending on Galactic latitude --
although this estimate probably suffers from stellar incompleteness
(Bailer-Jones \& Smith~\cite{cbj08a}).)  Galaxies will also be rare,
so in reality we should also modify their class fraction. Yet we
intentionally modify the class fractions just for a single class here
in order to better illustrate our prior modification method.

Results are presented in three ways. The first is a confusion matrix,
in which an object is assigned to the class with the largest DSC
output (most probable class) and the table shows the correct
classifications (on-diagonal values) and misclassification values
(off-diagonal values) as percentages.  Second, we show histograms of
the posterior probabilities. All histograms are density estimates with
unit area and always show the actual test data sets (no adjustment in
the number of plotted objects for the modified cases).  We then set
thresholds, $P_t$, on the posterior probabilities in order to build
samples (section~\ref{sample}). The third method of illustration shows how the
sample completeness and contamination varies with this threshold for
each output class.

\subsection{G=18.5 with all quasars in the training data}\label{g185}

We report only briefly on this case, primarily in order to motivate
the removal of the low EW quasars in experiments (2) and (3).

\begin{table} 
\begin{center}
\begin{minipage}{0.5\textwidth}
  \caption{Confusion matrix for class assignments from maximum
    probability. Each row corresponds to a true class and sums to
    100\%. Nominal priors,
    G=18.5\label{tab:nom185_allqso_contingency_table}}
  \vspace*{0.5em}
\begin{tabular}{lrrr}
\hline
         &  galaxy & quasar & star  \\
  GALAXY & 99.36   &  0.19  &  0.45 \\
  QUASAR &  1.03   & 96.04  &  2.94 \\
  STAR   &  0.49   &  0.88  & 98.63 \\
\hline
\end{tabular}
\end{minipage}
\end{center}
\end{table}

\begin{figure} 
\begin{center}
\includegraphics[width=0.46\textwidth]{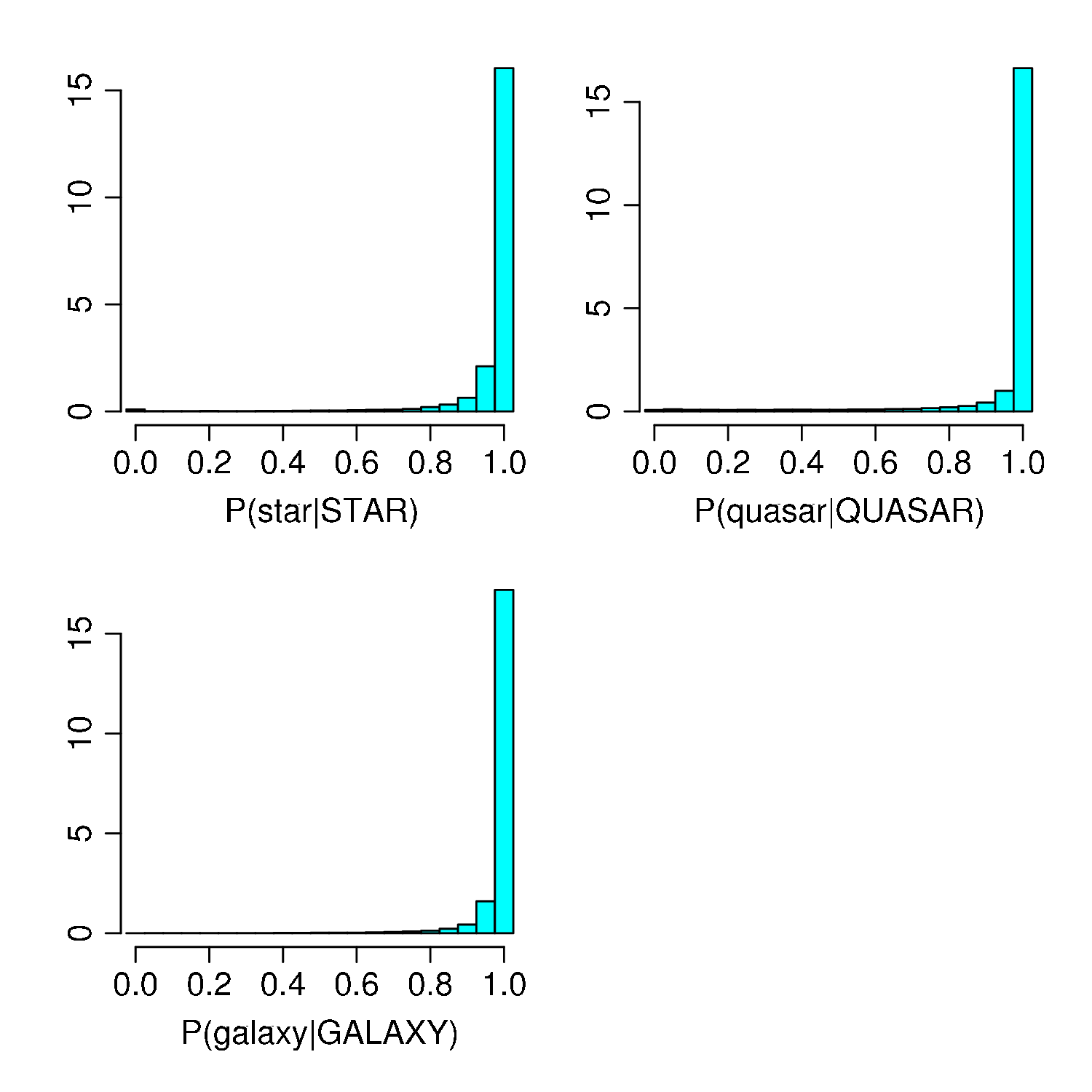}
\caption[Histograms $P({\tt class} | {\tt CLASS})$, nominal priors,
G=18.5]{Histograms of DSC outputs for each class showing how
  confident DSC is of identifying each class. Nominal
  priors, G=18.5
  \label{fig:nom185_allqso_correct_class_histogram}}
\end{center}
\end{figure}

\begin{figure} 
\begin{center}
\includegraphics[width=0.49\textwidth]{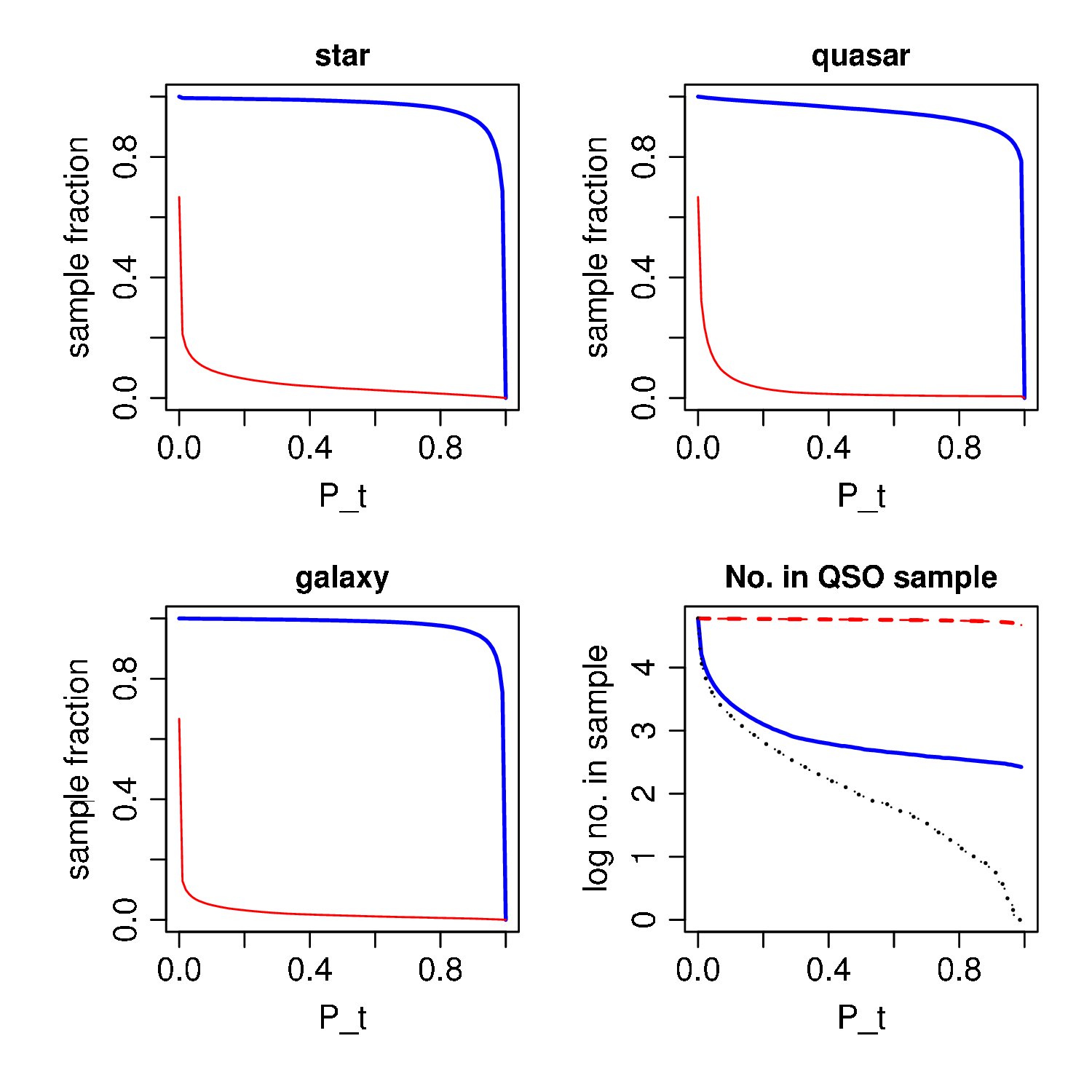}
\caption[Completeness \& contamination, nominal priors,
G=18.5]{Completeness (blue/thick line) and contamination (red/thin
  line) of a sample as a function of the probability threshold.  The
  bottom right panel shows the (logarithm) of the actual number of
  different types of class in the quasar sample: stars (blue/solid
  line); quasars (red/dashed lines); galaxies (black/dotted line).
  Nominal priors, G=18.5
  \label{fig:nom185_allqso_completeness_confusion}}
\end{center}
\end{figure}

\begin{figure} 
\begin{center}
\includegraphics[width=0.49\textwidth]{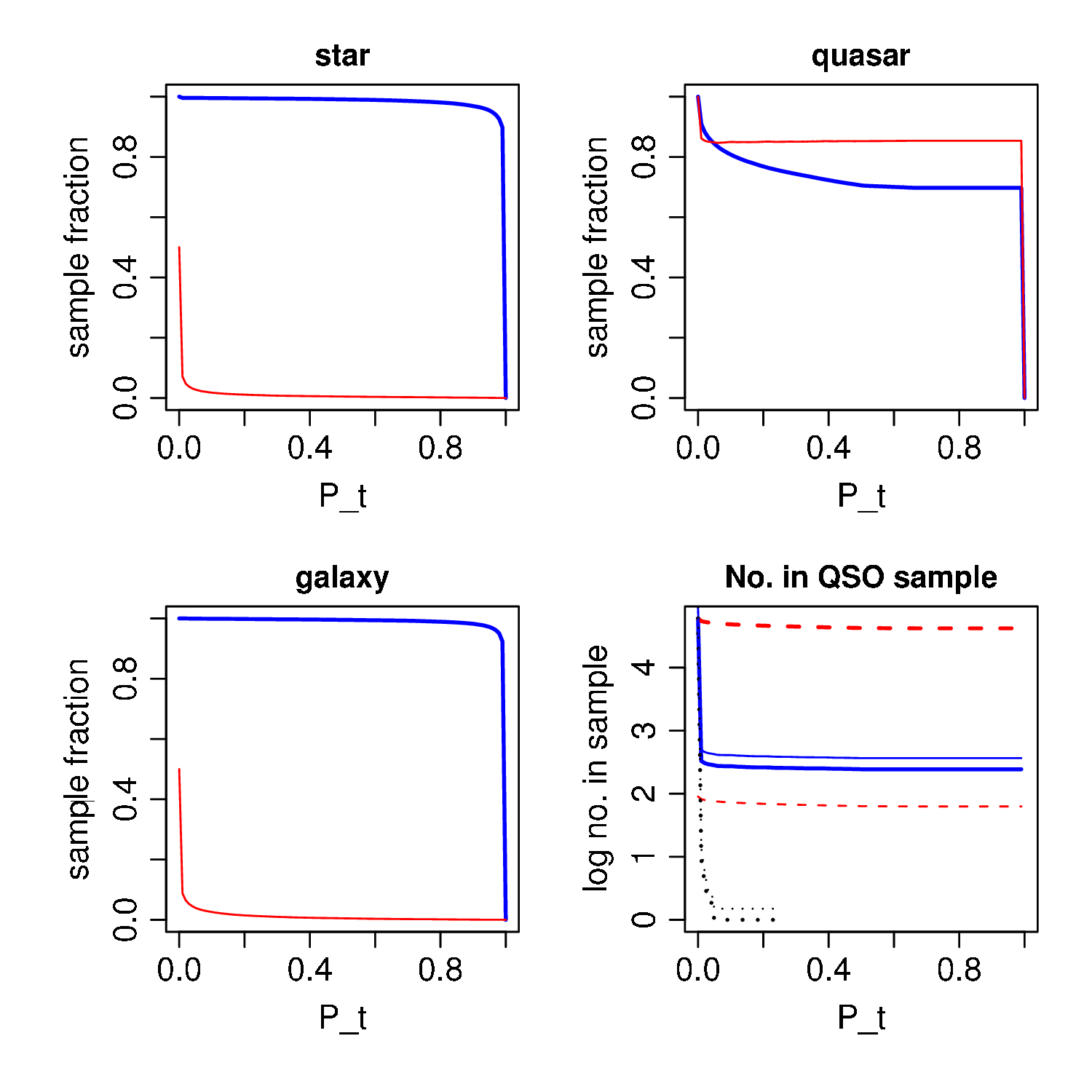}
\caption[Completeness \& contamination, modified priors,
G=18.5]{Completeness (blue/thick line) and contamination (red/thin
  line) of a sample as a function of the probability threshold.  The
  bottom right panel shows the (logarithm) of the actual number (thick
  lines) of different types of class in the quasar sample (i.e.\ in
  the test set): stars (blue/solid line); quasars (red/dashed
  lines). The thin lines show the corresponding {\em effective} number
  of objects for the modified case. Modified priors, G=18.5
\label{fig:mod185_allqso_completeness_confusion}}
\end{center}
\end{figure}

The confusion matrix (Table~\ref{tab:nom185_allqso_contingency_table})
shows that good classification accuracy is achieved with the nominal
model.  The histograms of the posterior probabilities
(Fig.~\ref{fig:nom185_allqso_correct_class_histogram}) further show
that these classifications are achieved with high confidence. This
translates into a satisfactory trade-off in sample completeness and
contamination at thresholds between 0.5 and 0.9
(Fig.~\ref{fig:nom185_allqso_completeness_confusion}).  \footnote{The
  completeness always decreases monotonically from 1 at $P_t=0$ (all
  objects included in sample) to 0 to $P_t=1$ (no objects in
  sample). The maximum contamination occurs at $P_t=0$ with a value
  depending on the class fractions (for equal class fractions in $K$
  classes it will be $1/K$) and drops to zero at $P_t=1$.}  The bottom
right-hand panel shows that stars rather than galaxies are the main
contaminant in the quasar sample. If we now proceed to the modified
model, the C\&C curves are very different
(Fig.~\ref{fig:mod185_allqso_completeness_confusion}).  While we would
expect higher contamination for a given completeness level for quasars
(because they are now very rare), we see that it is impossible to get
a low contamination quasar sample at all. The bottom right-hand panel
of that plot explains why: The effective number of stars in the sample
(366) is much higher than the effective number of quasars (63), so the
contamination is high, $366/(366+63) = 0.85$ (To recall: the effective
number is the number we {\em would} have in a data set which had the
modified class fractions.)

\begin{figure}
\begin{center}
\includegraphics[width=0.48\textwidth]{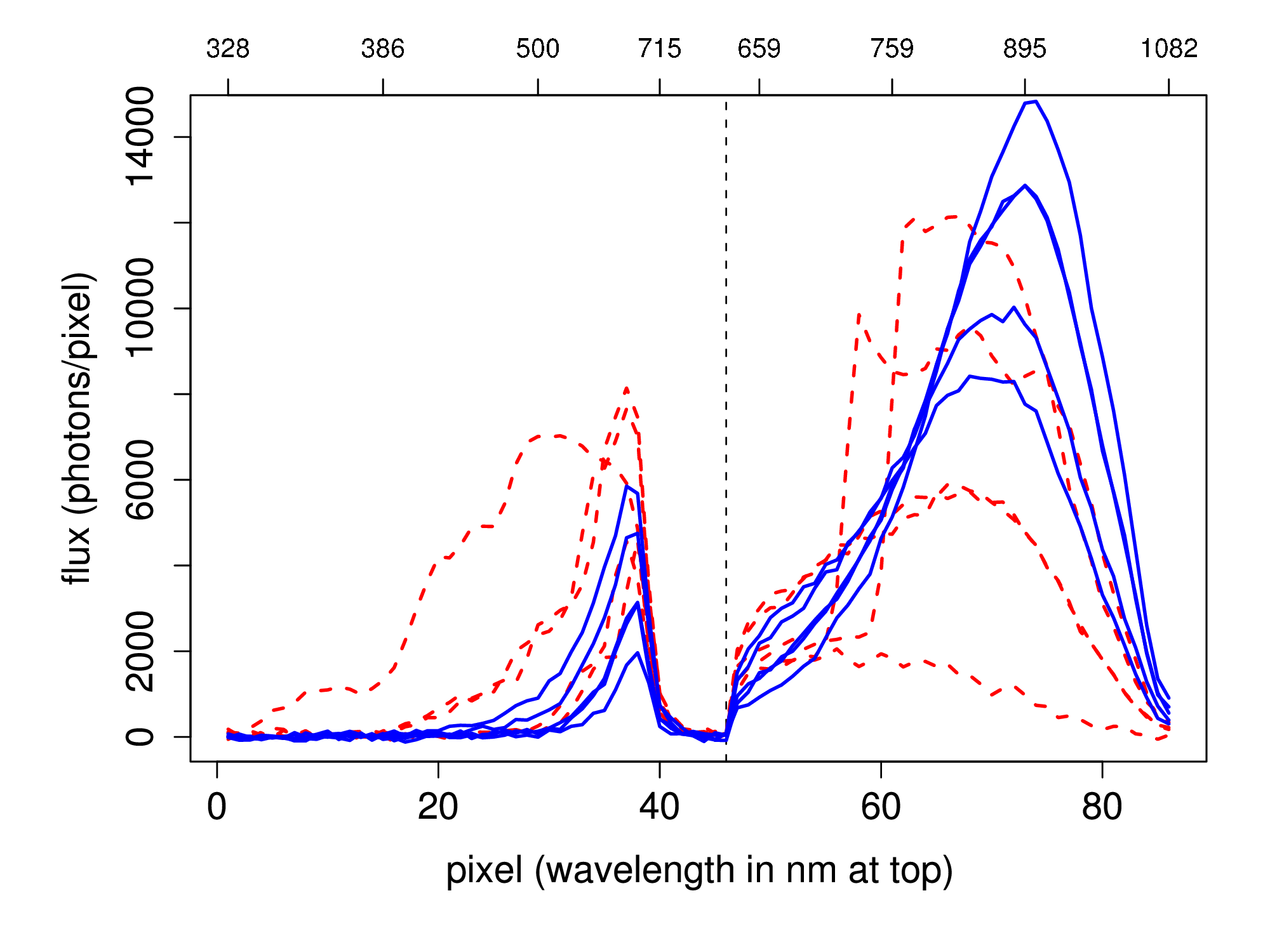}
\caption[]{Examples of low EW quasars (red/dashed lines) and those stars
  (blue/solid lines) which are the contaminants in the quasar sample
  in the modified G=18.5 model \label{fig:lowEWQSOs}}
\end{center}
\end{figure}

\begin{figure}
\begin{center}
\includegraphics[width=0.45\textwidth]{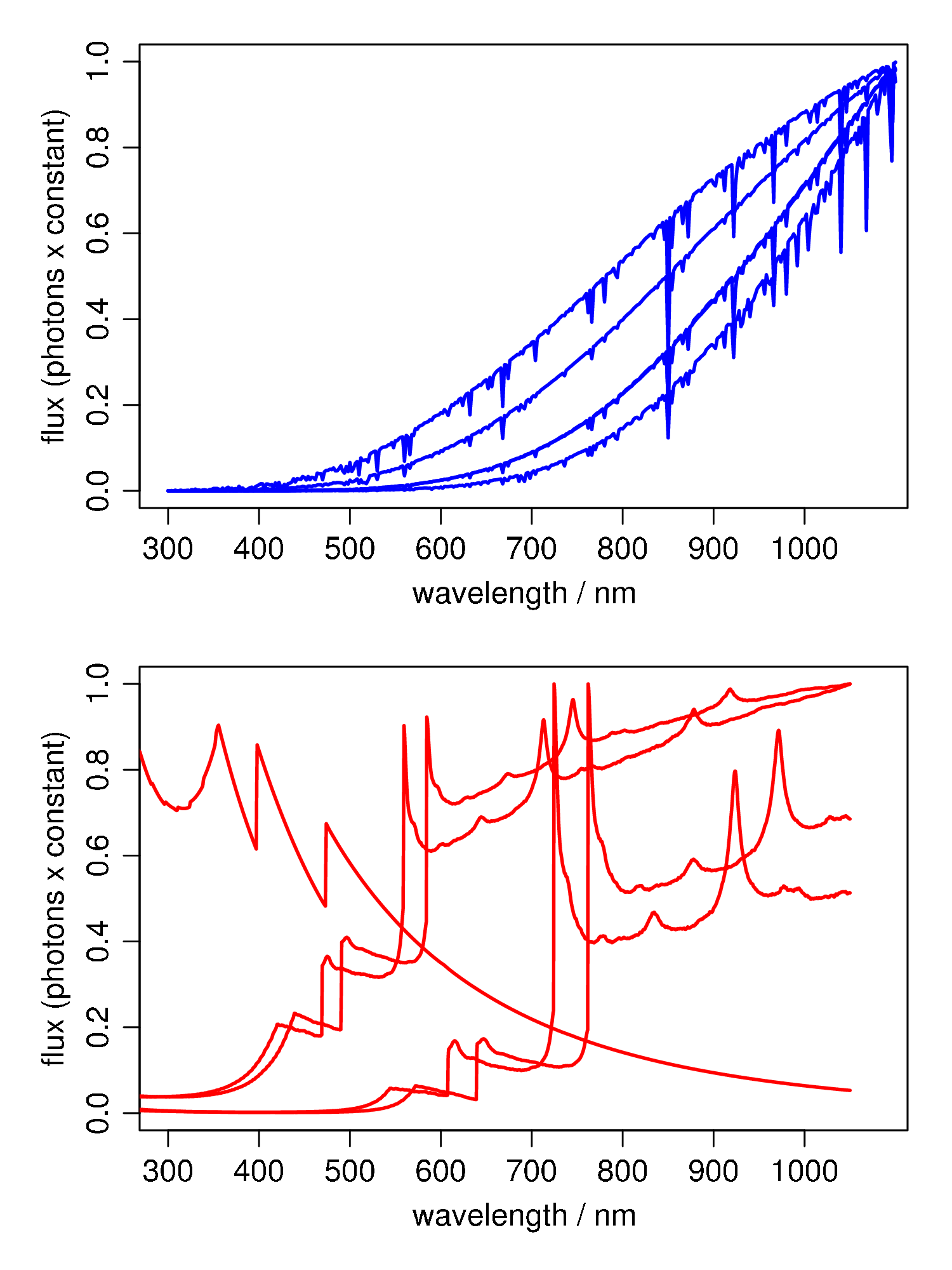}
\caption[Example stellar and quasar spectra]{The (noise-free) input spectra for the
  stars (top panel) and quasars (bottom panel) shown in
  Fig.~\ref{fig:lowEWQSOs}. The photon flux has been
  normalized in each case to have a maximum of
  1.0.\label{fig:lowEWQSOs_inputspectra}}
\end{center}
\end{figure}

Inspection of the contaminants shows that they are predominantly cool,
highly-reddened stars, with \teff\ between 4000 and 8000\,K and \av\
mostly between 8 and 10\,mag.  While the full stellar data set is
dominated by stars in this temperature range, the extinction
distribution peaks towards lower extinctions
(Fig.~\ref{large_star_aps}), so this tendency for highly extincted
contaminants is significant.  Figs.~\ref{fig:lowEWQSOs}
and~\ref{fig:lowEWQSOs_inputspectra} plot a random selection of these
stellar contaminants along with a random selection of quasars with
lower values of the emission line EW parameter.  Note how the very
broad PSF of BP/RP is smoothing out the sharp features in the quasar
spectra, especially at the red of the two channels where the
dispersion is lower. The narrower quasar emission lines are not
resolved, and these objects look more like the smoother stellar
spectra.  We hypothesize that if we remove such quasars from our
training set, and thus from our definition of quasars, then the SVM
should not so readily confuse these stars with our quasar class. We
test this in the next experiment (section~\ref{g185nolowEWQSO}).

\begin{table} 
\begin{center}
\begin{minipage}{0.5\textwidth}
  \caption{Model-based priors for the nominal, $P(C_j |
    \theta^{nom})$, and modified, $P(C_j | \theta^{mod})$, cases for
    the full training data and the case in which low EW quasars have been
    removed (``nlEW''). For comparison we show the class fractions relevant to he nominal
    models, $f^{train}_i$, and the modified models
    $f^{mod}_i$\label{tab:mbp}}
\vspace*{0.5em}
\begin{tabular}{llllll}
                         & data & $G$   & star   & quasar  &  galaxy \\
\hline
$P(C_j | \theta^{nom})$   & full & 18.5 & 0.3380 & 0.3279 & 0.3341 \\
$f^{train}_i$             & full & 18.5 & 0.3333 & 0.3333 & 0.3333 \\
$P(C_j | \theta^{mod})$   & full & 18.5 & 0.4965 & 0.002514 & 0.5010 \\
$f^{mod}_i$               & full & 18.5 & 0.4998 & 0.000500 & 0.4998 \\
\hline
$P(C_j | \theta^{nom})$   & nlEW & 18.5 & 0.367 & 0.283 & 0.350 \\
$P(C_j | \theta^{nom})$   & nlEW & 20.0 & 0.368 & 0.260 & 0.372 \\
$f^{train}_i$             & nlEW & both & 0.388 & 0.225 &  0.388 \\
$P(C_j | \theta^{mod})$   & nlEW & 18.5 & 0.4983 & 0.000328 & 0.5013 \\
$P(C_j | \theta^{mod})$   & nlEW & 20.0 & 0.4762 & 0.000277 & 0.5234 \\
$f^{mod}_i$               & nlEW & both & 0.4998 & 0.000500 & 0.4998 \\
\hline
\end{tabular}
\end{minipage}
\end{center}
\end{table}

Table~\ref{tab:mbp} lists the model-based priors (section~\ref{mbp}).
The first line is for the nominal model. The second row gives, for
comparison, the fraction of objects in each true class, $i$, in the
training data. These we may consider as frequentist estimates of the
model priors, insofar as the frequency distribution of the classes
dicates these. At least for this SVM model with equal class fractions,
the model-based priors are close to the class fractions.

The third and fourth lines give the same but for the modified
model. Now we see that the modified class fraction, $f^{mod}_i$, for
the quasars is not a good proxy for the model-based prior. This
implies that its use in equation~\ref{eqn:modprob} will give poor
estimates for the true posterior probabilities. We could attempt to
improve this by an iterative procedure: Now that we have the
model-based priors, we can recalculate the model posteriors directly
from Bayes' equation (\ref{bayes}) -- rather than our approximation
(equation~\ref{eqn:modprob}) -- and then recalculate the model-based
priors with equation~\ref{margin}. However, we don't do this because
in our main experiments (next), the discrepancy is not as large.

\subsection{G=18.5 with low EW quasars removed from the training data}\label{g185nolowEWQSO}

Motivated by the results of the previous experiment, we removed the
low equivalent width quasars (EW$<5000$\,\AA) from the training data
set (2099 of 5000) and re-tuned and re-trained the SVM. (The choice of
$5000$\,\AA\ is somewhat arbitrary.) The test set is unchanged.

\subsubsection{The nominal model}

\begin{table} 
\begin{center}
\begin{minipage}{0.5\textwidth}
  \caption{Confusion matrix for class assignments from maximum
    probability. Each row corresponds to a true class and
    sums to 100\%. Nominal priors, G=18.5, no low EW quasars in
    training data\label{tab:nom185_contingency_table}}
\vspace*{0.5em}
\begin{tabular}{lrrr}
\hline
         &  galaxy & quasar & star  \\
  GALAXY & 99.37   &  0.00  &  0.63 \\
  QUASAR &  4.22   & 85.59  & 10.19 \\
  STAR   &  0.68   &  0.13  & 99.19 \\
\hline
\end{tabular}
\end{minipage}
\end{center}
\end{table}

Comparing the confusion matrix
(Table~\ref{tab:nom185_contingency_table}) to that in the
previous experiment, we now see that fewer quasars are correctly
classified, with 10\% being misclassified as stars. Yet this loss of
quasars is balanced by the fact that six times fewer stars are now
misclassified as quasars (0.13\% rather than 0.88\% previously,
or 78 stars rather than 528). This is what we wanted to achieve by
modifying the training sample. 

Note how few galaxies are misclassified as stars and quasars, and how
this has hardly changed from the previous experiment. In all
experiements we have performed with these data (including many not
reported here), the galaxies are always classified with high
confidence. As we are not changing the class fractions for galaxies --
they are acting mostly to make the classification problem for the SVM
harder -- we focus on the stars and quasars from now on.

\begin{figure*} 
\begin{center}
\includegraphics[width=0.71\textwidth]{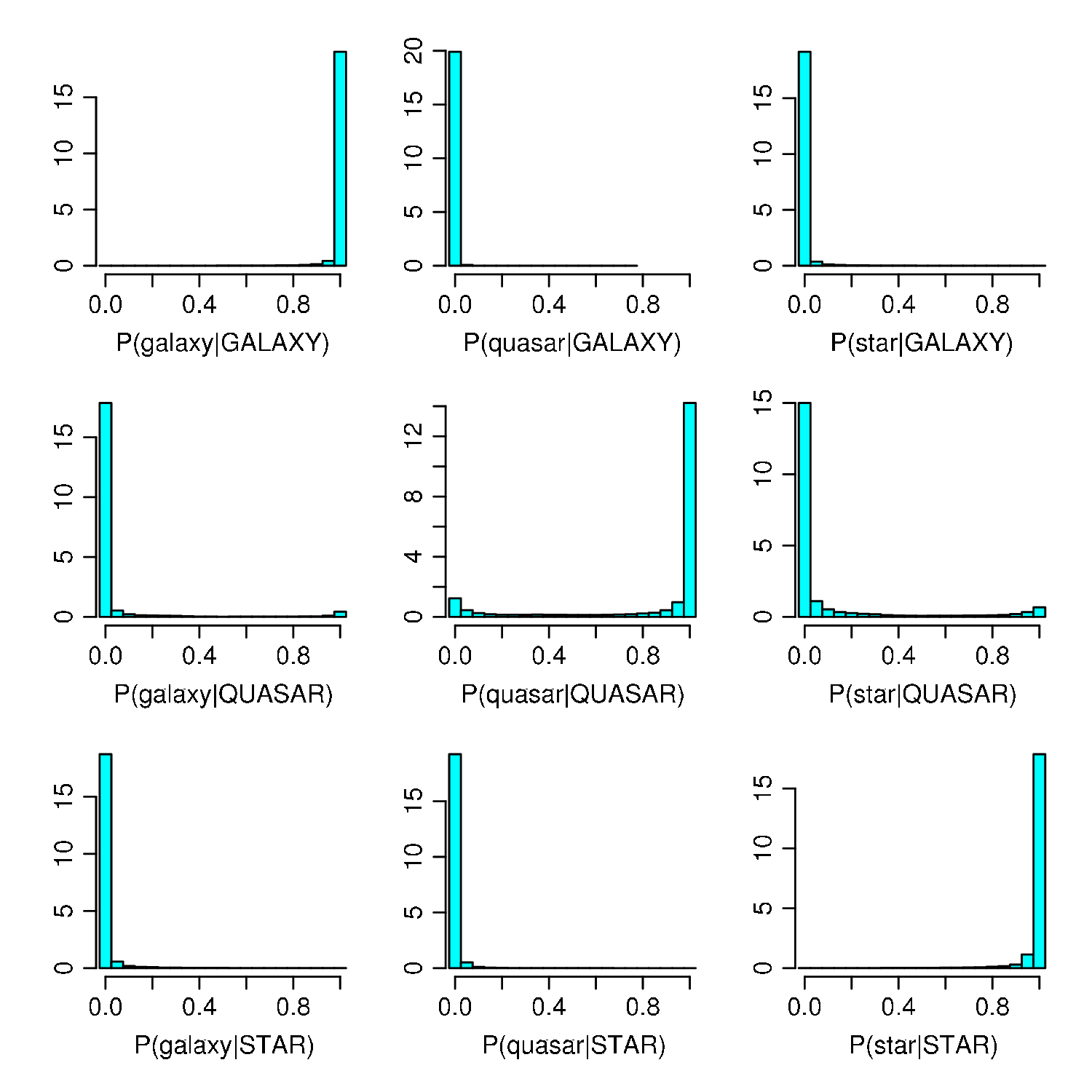}
\caption[Histograms $P({\tt class} | {\tt CLASS})$, nominal priors,
G=18.5]{Histograms of the DSC class posterior probabilities, shown for each output class
  (columns) split for objects of each true class (rows). Nominal
  priors, G=18.5, no low EW quasars in training data
  \label{fig:nom185_all_histogram}}
\end{center}
\end{figure*}

\begin{figure} 
\begin{center}
\includegraphics[width=0.49\textwidth]{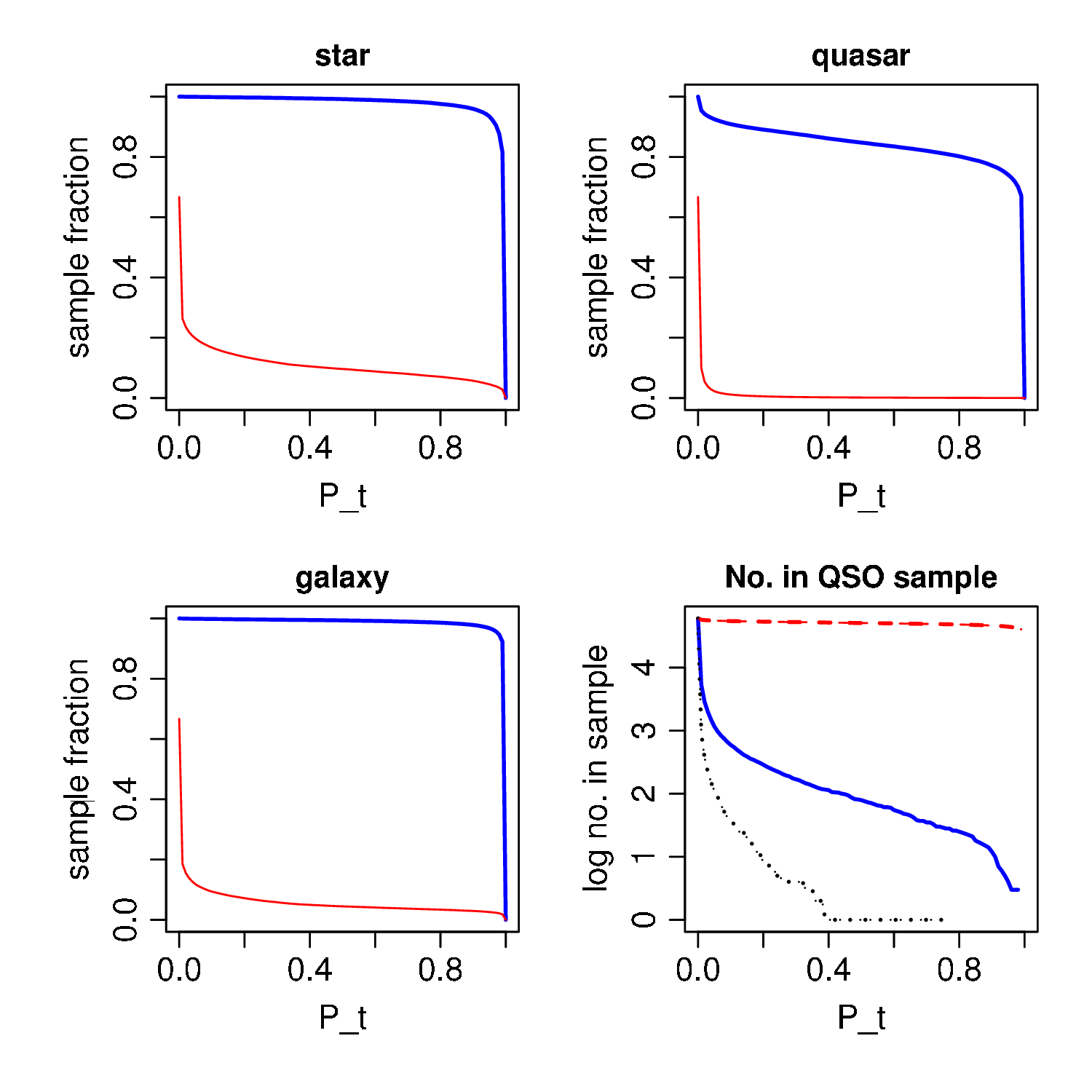}
\caption[Completeness \& contamination, nominal priors,
G=18.5]{Completeness (blue/thick line) and contamination (red/thin
  line) of a sample as a function of the probability threshold.  The
  bottom right panel shows the (logarithm) of the actual number of
  different types of class in the quasar sample: stars (blue/solid
  line); quasars (red/dashed lines); galaxies (black/dotted line).
  Nominal priors, G=18.5, no low EW quasars in training data
\label{fig:nom185_completeness_confusion}}
\end{center}
\end{figure}

We summarze the model confidence (posterior probabilties) in the
histograms in Fig.~\ref{fig:nom185_all_histogram}. We can read several
things from this: the leading diagonal shows $P({\tt class} |
{\tt CLASS})$, how confident the true positives are; the central
row shows $P({\tt class} | {\tt QUASAR})$, the probabilities assigned
to each class for true quasars; the central column shows $P({\tt
  quasar} | {\tt CLASS})$, the quasar probabilities assigned to
objects of each true class. We see that the confidences for the
correct classes are still high.  However, comparing with the results
of the previous experiment
(Fig.~\ref{fig:nom185_allqso_correct_class_histogram}), we now see a
set of true quasars which are now assigned very low probabilities,
$P({\tt quasar} | {\tt QUASAR})$.  These are predominantly (but not
exclusively) the low EW quasars that were removed from the training
data.  These low probability cases are reflected as a dip in the
quasar completeness curve
(Fig.~\ref{fig:nom185_completeness_confusion}) at low $P_t$. These
quasars show up in the star sample and we correspondingly see a higher
contamination for the stars.  Yet the positive trade off from this is
a lower quasar sample contamination.  To give numbers: At $P_t=0.8$
the completeness and contamination fractions for stars are 0.976 and
0.071 respectively and for quasars are 0.8020 and 0.0005. So we see
that even with the nominal model we can get a very pure quasar sample,
but at the cost of a rather unpure stellar one.

\subsubsection{The modified model}\label{modmodel}

\begin{figure*} 
\begin{center}
\includegraphics[width=0.71\textwidth]{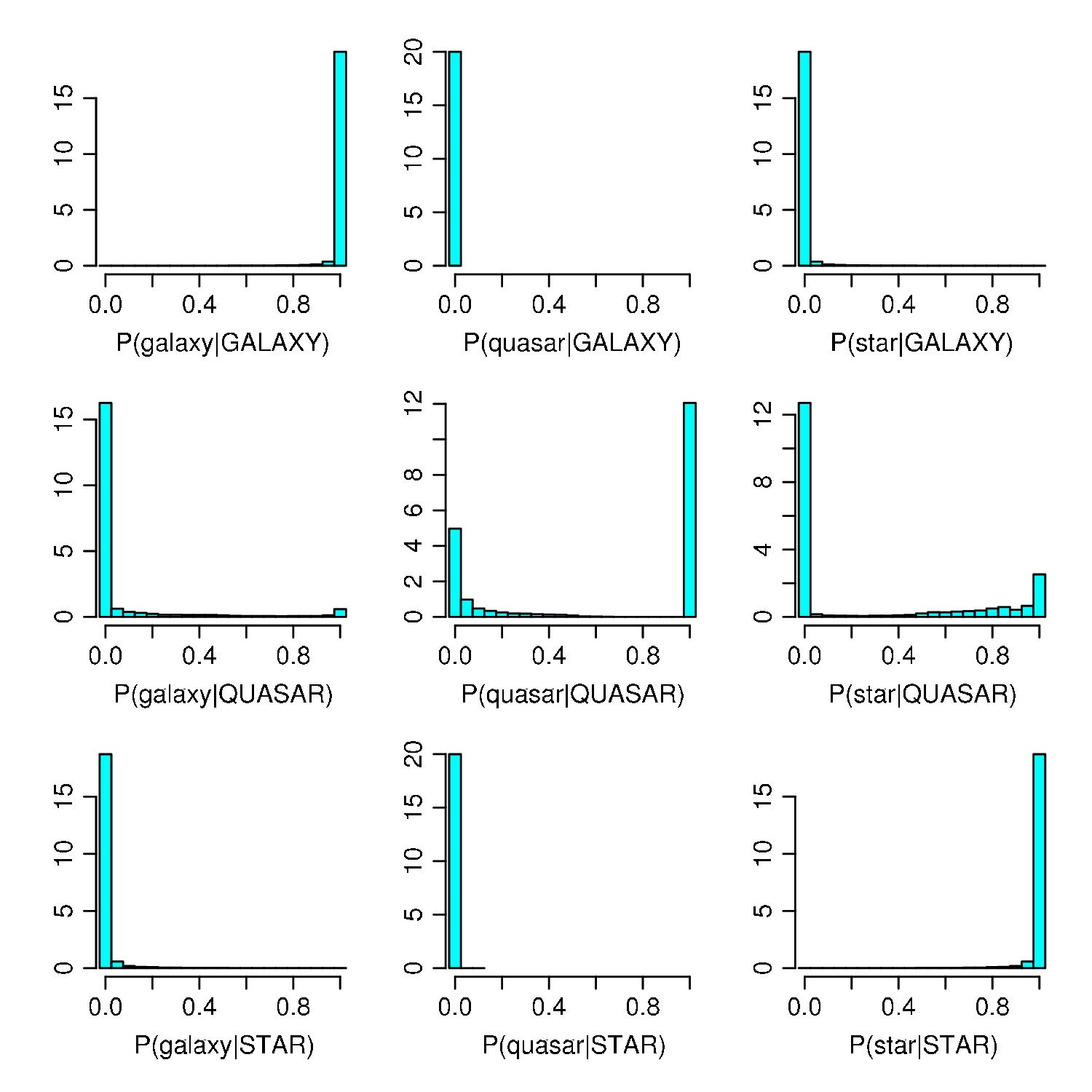}
\caption[Histograms $P({\tt class} | {\tt CLASS})$, modified priors,
G=18.5]{Histograms of the DSC class posterior probabilities, shown for each output class
  (columns) split for objects of each true class (rows). Modified
  priors, G=18.5, no low EW quasars in training data
  \label{fig:mod185_all_histogram}}
\end{center}
\end{figure*}

\begin{figure} 
\begin{center}
\includegraphics[width=0.49\textwidth]{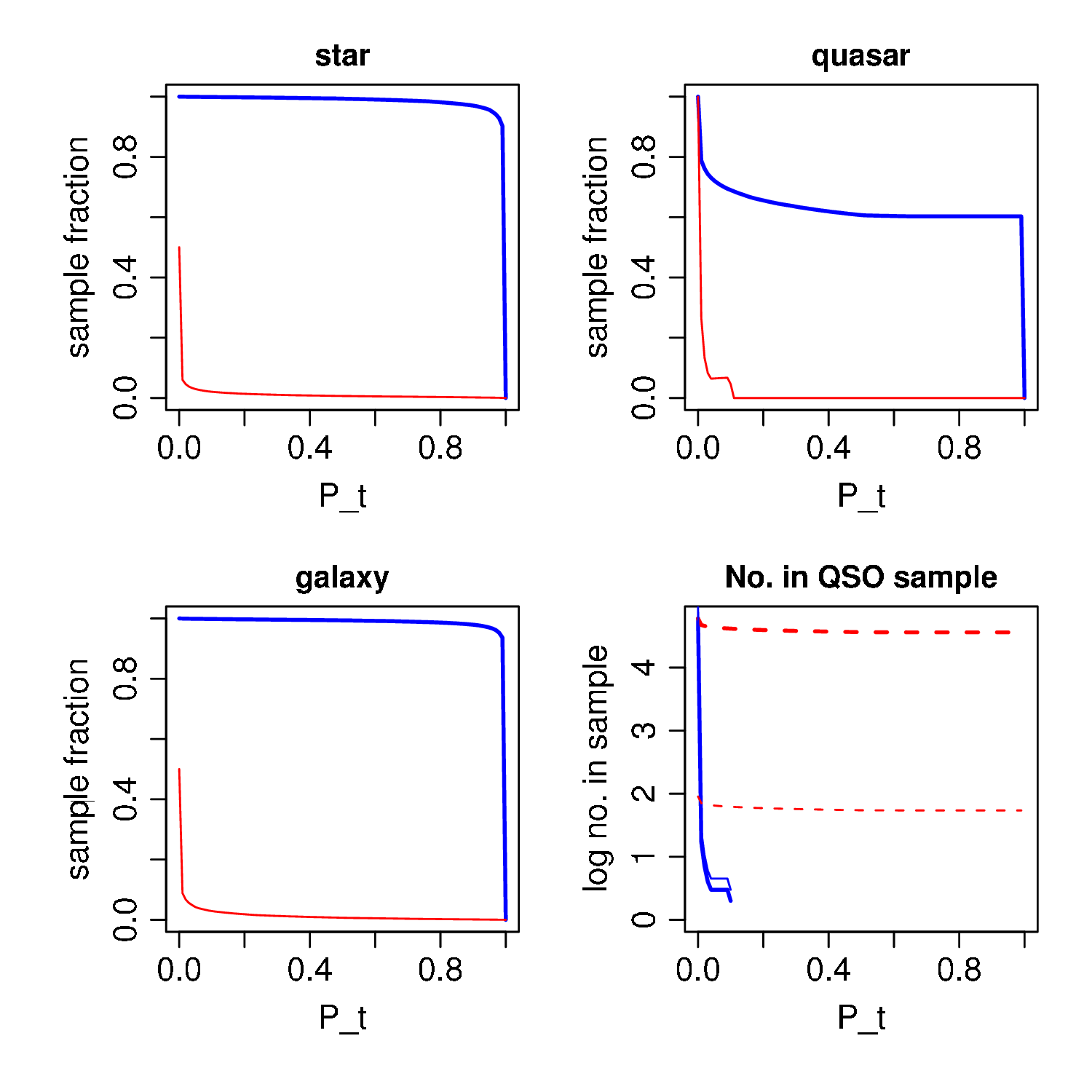}
\caption[Completeness \& contamination, modified priors,
G=18.5]{Completeness (blue/thick line) and contamination (red/thin line)
of a sample as a function of the probability threshold.
The bottom right panel shows the (logarithm) of the actual number (thick lines) of different types of class in the quasar sample (i.e.\ in the test set): stars (blue/solid line); quasars (red/dashed lines). The thin lines show the corresponding {\em effective} number of objects for the modified case. There are no galaxy contaminants.
Modified priors, G=18.5, no low EW quasars in training data
\label{fig:mod185_completeness_confusion}}
\end{center}
\end{figure}

We turn now to the modified model in which our prior is that quasars
are 1000 times rarer. The histograms of the posteriors are show in
Fig.~\ref{fig:mod185_all_histogram} which we may compared to the nominal
case in Fig.~\ref{fig:nom185_all_histogram}.  The main difference can
be seen in the central plot, where we now observe a significant peak
around zero probability for $P({\tt quasar} | {\tt QUASAR})$: the
modified model has a high barrier to classifying anything as a
quasar. This is the result of the probability remapping,
equation~\ref{eqn:modprob} (whereby the impact of the normalization
should not be ignored).  This provides more discrimination between
those quasars which, in the nominal case, were all assiged an output
probability of almost 1.0.

Fig.~\ref{fig:mod185_completeness_confusion} shows the C\&C for the
modified model. It is significantly different from the nominal
case. The quasar completeness is slightly reduced, but we win a much
lower contamination. Indeed, at $P_t=0.11$ the contamination drops to
zero for a sample completeness of 65\%.  This translates to a
contamination of less than 1 in $0.65 \times 60 000 =$\,39\,000. Not
only is this better than for the nominal model (where the
contamination only drops to zero at $P_t=1.0$), but now the star and
galaxy contaminations are also much lower.

The bottom right-hand plot of
Fig.~\ref{fig:mod185_completeness_confusion} shows the actual (solid
line) and effective (thin line) number of objects in the quasar
sample.  At $P_t=0.8$ these are 36149 ($=10^{4.56}$) and 54
($=10^{1.73}$) respectively. If we had done this experiment with a
test set containing, say, only 2000 quasars, then the effective number
of quasars in the resultant sample would only have been $54 \times
(2000/60\,000) = 1.8$.  This is not enough to draw results of any
significance and explains why we need large test sets for evaluating
the modified model.

\begin{table} 
\begin{center}
\begin{minipage}{0.5\textwidth}
  \caption{Confusion matrix for class assignments achieved by applying
    a probability threshold. Each row gives the percentage of objects of
    each true class assigned to the different output classes
    (columns).  As an object may now be assigned to more than one
    class, the values in a row no longer sum to 100\%, plus some
    objects may remain unclassified. The thresholds applied are $P_t = 0.2$
    for quasars and $P_t=0.8$ for stars and galaxies. Modified
    priors, G=18.5\label{tab:mod185_contingency_table}}
  \vspace*{0.5em}
\begin{tabular}{lrrrrl}
\hline
         &  galaxy  & quasar  & star  & unclassified & Effective \\
         &          &         &       &              & fraction \\
GALAXY   &  98.97   &   0.00  & 0.64  &    0.73      & 1.0      \\
QUASAR   &   6.82   &  62.00  & 26.37 &    8.73      & 0.001    \\
STAR     &   0.78   &   0.00  & 98.69 &    1.09      & 1.0      \\
\hline
\end{tabular}
\end{minipage}
\end{center}
\end{table}

The confusion matrix for this modified case is shown in
Table~\ref{tab:mod185_contingency_table}.  As class assignments are
now based on thresholds rather than maximum probability, and because we can set these thresholds
independently of one another, a given object may be assigned to more
than one class or it may remain unclassified.  Therefore the values in
a row no longer sum to 100\%. Here we use thresholds of $0.8, 0.2,
0.8$ for galaxy, quasar and star respectively.  The quasar threshold
is chosen to give zero quasar contamination and delivers a good
completeness of 62\%. The star and galaxy threholds were also chosen
from inspecting the C\&C curves in
Fig.~\ref{fig:mod185_completeness_confusion}, and yield high
completeness (around 99\%) and about 0.7\% cross contamination. From a
first glance at the table we might think that this star sample is heavily
contaminated by quasars (26.37\%). Yet we must remember that quasars are
effectively rare, so the fraction of quasar contaminants expressed as
percentage of objects in the star output is
\[
100\% \times \frac{26.37 \times 0.001}{26.37 \times 0.001 + 98.69 \times 1.0 + 0.64 \times 1.0} =
0.03\%
\]
(The quasar contamination of the galaxy sample is even
smaller). Recall also that the histograms, such as
  Fig.~\ref{fig:mod185_all_histogram}, show the actual number
  of objects in the test set, not the effective number. So what looks
  like a relatively large number of quasars confidently classified as
  stars corresponds to a far smaller number in the target population.


\begin{figure} 
\begin{center}
\includegraphics[width=0.48\textwidth]{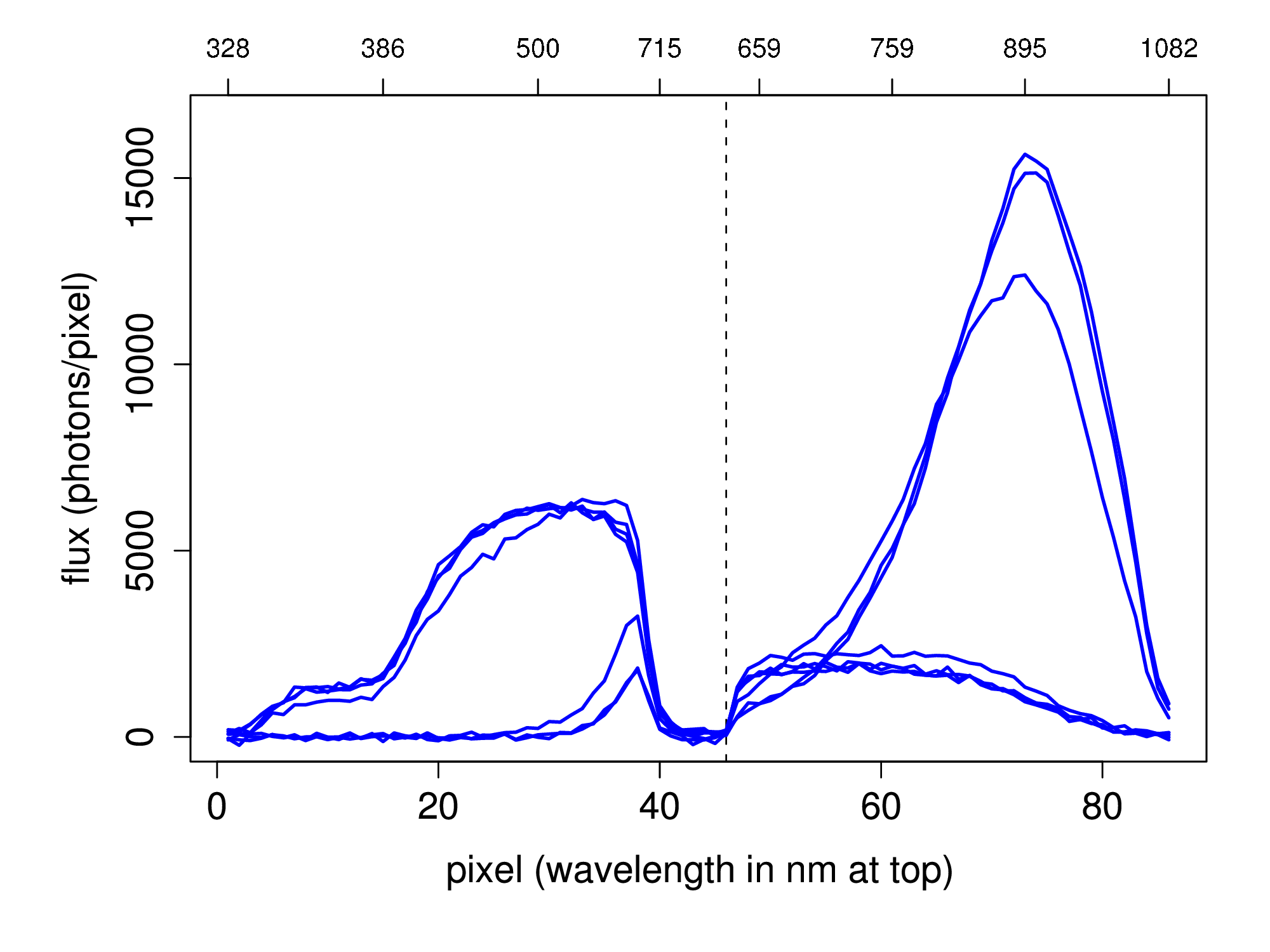}
\caption{Spectra of stars which are most confused as quasars, namely those 7 stars which have the highest values of $P^{mod}({\tt quasar} | {\tt STAR})$. G=18.5, modified priors, no low EW quasars in training data.
  \label{fig:mod_g185_stars_confused_as_quasars}}
\end{center}
\end{figure}

\begin{figure}
\begin{center}
\includegraphics[width=0.45\textwidth]{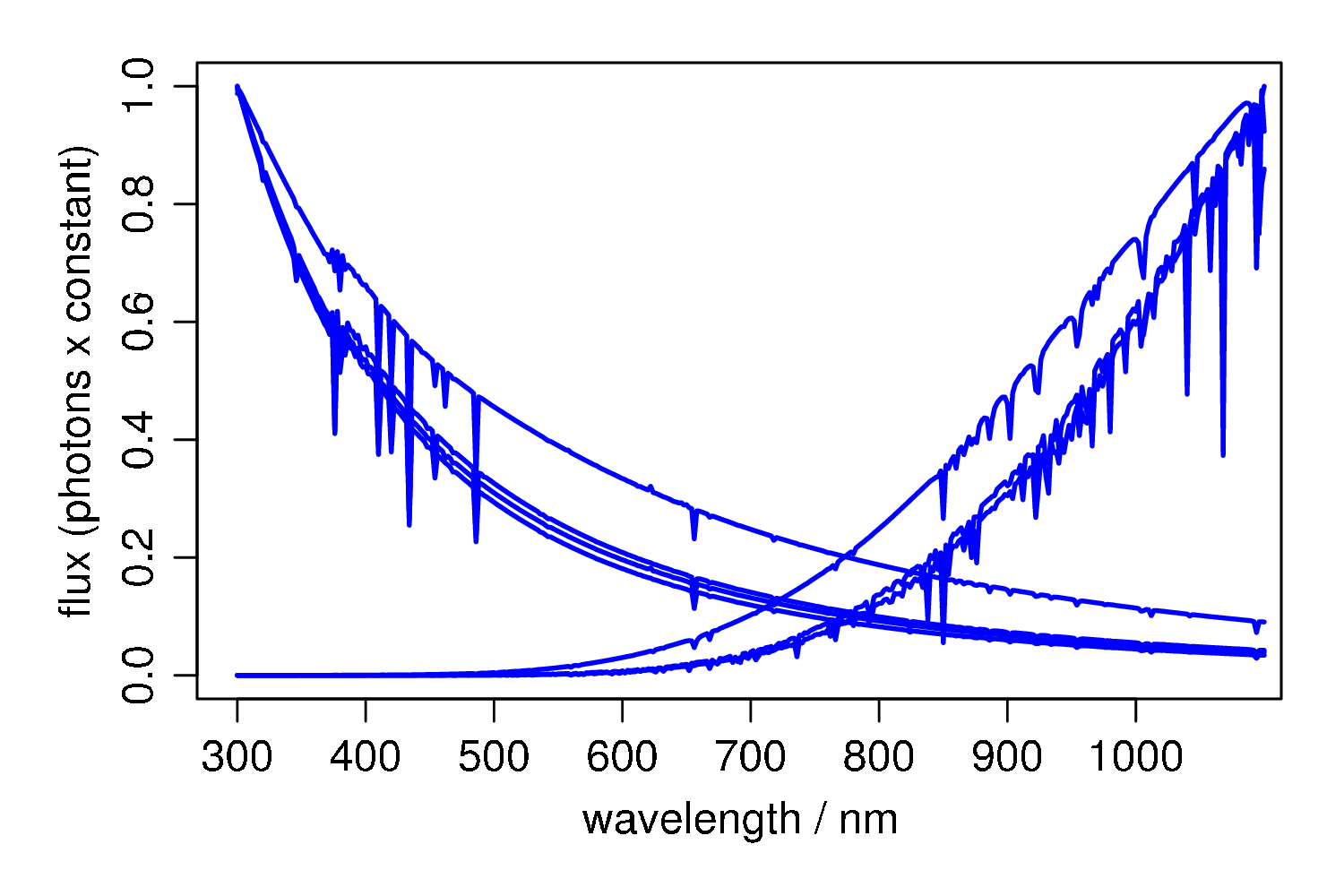}
\caption[]{The (noise-free) input spectra for the stars shown in
  Fig.~\ref{fig:mod_g185_stars_confused_as_quasars}. The photon flux
  has been normalized in each case to have a maximum of
  1.0.\label{fig:mod_g185_stars_confused_as_quasars_inputspectra}}
\end{center}
\end{figure}

It is again interesting to identify the misclassifications.  Those
true stars which are most confused as quasars are show in
Fig.~\ref{fig:mod_g185_stars_confused_as_quasars}. As can be seen from
the plot of the corresponding library spectra
(Fig.~\ref{fig:mod_g185_stars_confused_as_quasars_inputspectra}), four
of these are hot stars with \teff\,$\simeq$\,40\,000\,K. The other
three are cooler (two around 4000\,K, one around 7000\,K)
but have very high extinctions (8--9
magnitudes \av). This gives rise to the much larger flux in the red
which DSC confuses with a quasar-like broad emission line feature.


We recall that in this experiment we removed the low EW quasars from the
training set, but not the test set. We may, therefore, expect this
model to perform particularly badly on these objects. However, if we
plot Fig.~\ref{fig:mod185_all_histogram} just for these
objects it turns out to be broadly similar, although with about half
as many objects in the $P({\tt quasar} | {\tt low EW QUASAR}) = 1$ bin
and twice as many in the $P({\tt quasar} | {\tt low EW QUASAR}) = 0$
bin. So while this model doesn't do as well on the objects omitted
from the training data, it is still able to assign many a high
probability. That is, DSC has been able to extend its recognition of
quasars to lower EW cases than it was trained on.

\subsubsection{Testing the predictions of the modified model}\label{testpred}

\begin{figure} 
\begin{center}
  \includegraphics[width=0.30\textwidth]{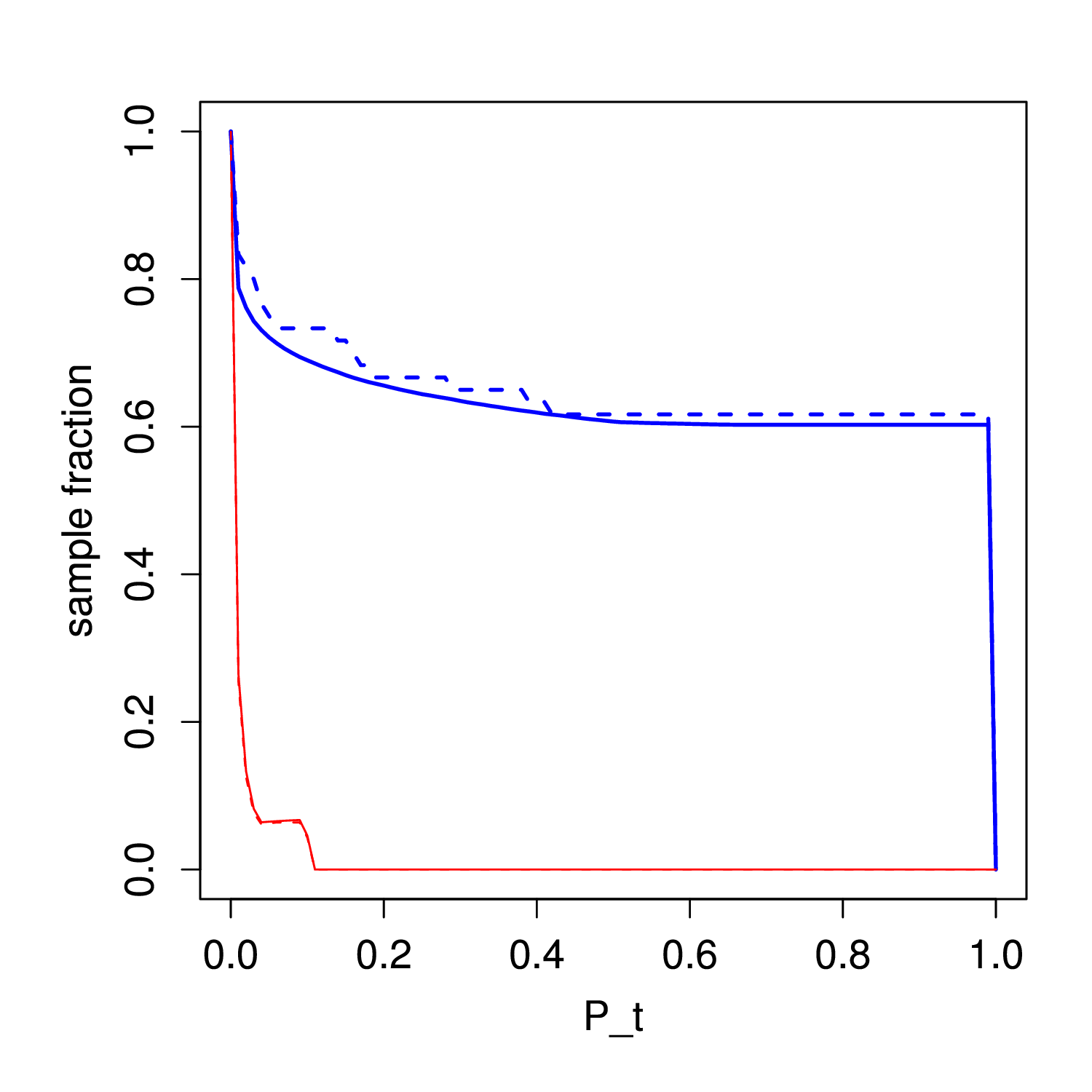}
  \caption[Comparison of predicted and measured quasar
  C\&C]{Comparison of the completeness (blue/thick lines) and
    contamination (red/thin lines) obtained with the modified model at
    G=18.5 (no low EW quasars in training data). The solid lines are the
    predictions (same as those in the the top right panel of
    Fig.~\ref{fig:mod185_completeness_confusion}) and the dashed lines
    are the measurements from a data set in which quasars really
    are rare (class fractions of (1,1/1000,1))}
\label{fig:mod_cc_comparison}
\end{center}
\end{figure}

The C\&C plots for the modified model are of course just predictions
of what the C\&C would be {\em if} we applied the modified model to a
real data set which has the modified population fractions.  We can
test these predictions by building a new test data set in which the
population fractions really are (1,1/1000,1).  To do this we took the
original test data set but retained just a random selection of 1/1000
of the quasars (i.e.\ 60 quasars). We applied the modified model and
calculated the C\&C (now using equation~\ref{eqn:compcont} of course,
not equation~\ref{eqn:mod_compcont}).  As there are only 60 quasars in
this set, the predictions have high variance, so we bootstrapped the
selection 20 times.  Fig.~\ref{fig:mod_cc_comparison} shows that the
measured C\&C are very close to the predictions, thus validating our
approach.

\subsubsection{The nominal model yields high contamination when quasars are rare}\label{nom_limited}

\begin{figure} 
\begin{center}
  \includegraphics[width=0.30\textwidth]{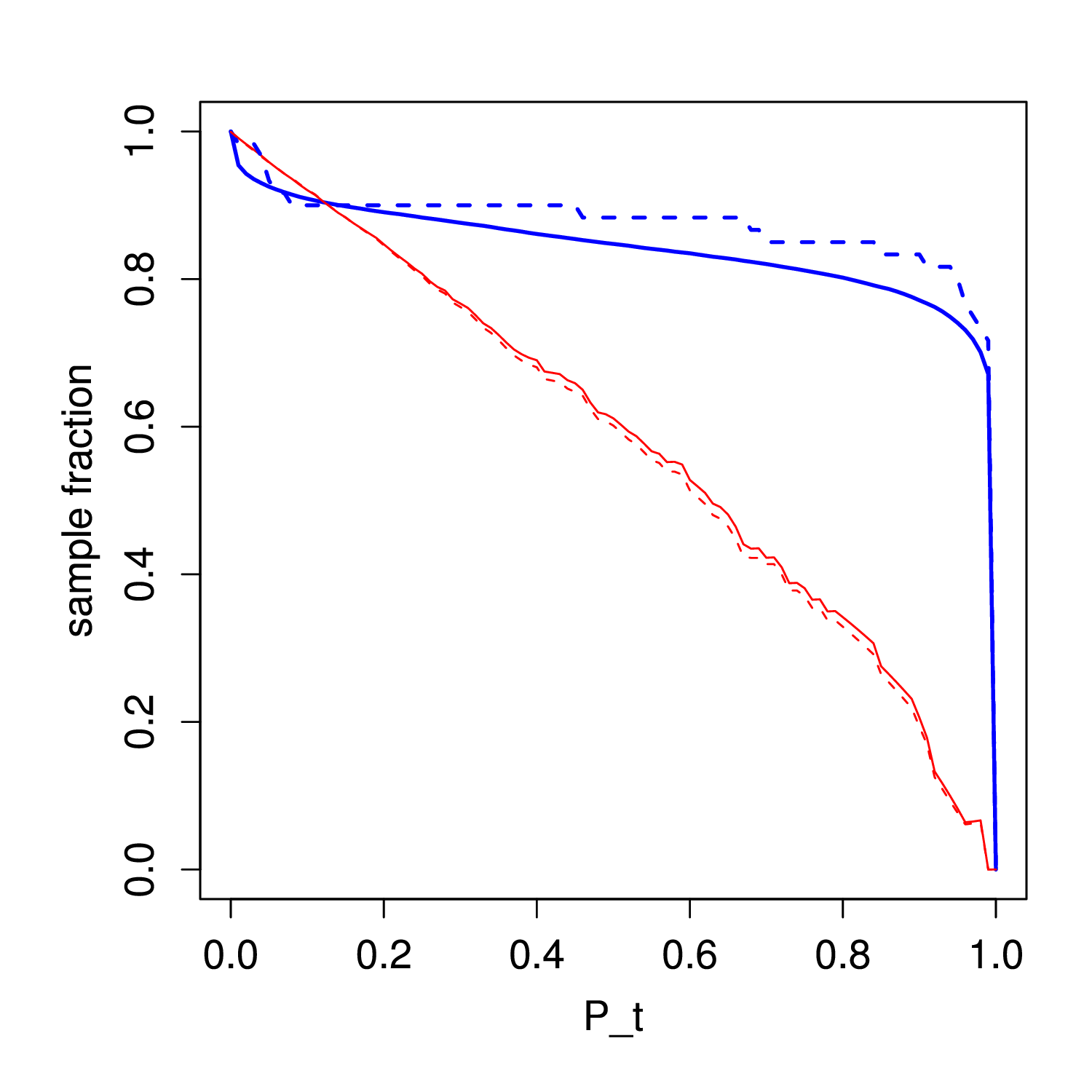}
  \caption[Comparison of predicted and measured quasar
  C\&C]{As Fig.~\ref{fig:mod_cc_comparison} but now for the nominal model}
\label{fig:nom_cc_comparison}
\end{center}
\end{figure}

Theoretical arguments aside, could we nonetheless use the nominal
model to classify data sets in which quasars are rare? To assess this,
we applied the nominal model to the few-quasar data set from the above
test (section~\ref{testpred}).  The measured C\&C are plotted in
Fig.~\ref{fig:nom_cc_comparison} as the dashed lines. These agree well
with the C\&C we would predict on the effective test data set (i.e.\
using equation~\ref{eqn:mod_compcont}) but using the nominal model,
which are shown as the solid lines in
Fig.~\ref{fig:nom_cc_comparison}.  While they agree, this figure shows
that the sample contamination is far higher when using the nominal
model than the modified model (Cf.\
Fig.~\ref{fig:mod_cc_comparison}). That is, the modified model is
superior at obtaining pure quasar samples. This demonstrates that we
must take into account the expected class fractions (i.e.\ use a
suitable prior) when applying a classifier to a data set in which we
expect a class to be rare.
\newline

\noindent
Finally for this experiment, we note that the model-based priors agree quite well with the
training data class fraction or the modified class fraction for the
nominal and modified models respectively (Table~\ref{tab:mbp}).

\subsection{G=20.0 with low EW quasars removed from the training data}\label{g200nolowEWQSO}

\begin{table} 
\begin{center}
\begin{minipage}{0.5\textwidth}
\caption{Confusion matrix for class assignments from maximum
    probability. Each row corresponds to a true class and
    sums to 100\%. Nominal priors, G=20.0, no low EW quasars in training data
\label{tab:nom200_contingency_table}}
\vspace*{0.5em}
\begin{tabular}{lrrr}
\hline
         &  galaxy & quasar & star  \\
  GALAXY & 93.53   &  0.12  &  6.66 \\
  QUASAR &  7.40   & 77.13  & 15.47 \\
  STAR   & 13.92   &  0.21  & 85.87 \\
\hline
\end{tabular}
\end{minipage}
\end{center}
\end{table}

\begin{figure} 
\begin{center}
\includegraphics[width=0.46\textwidth]{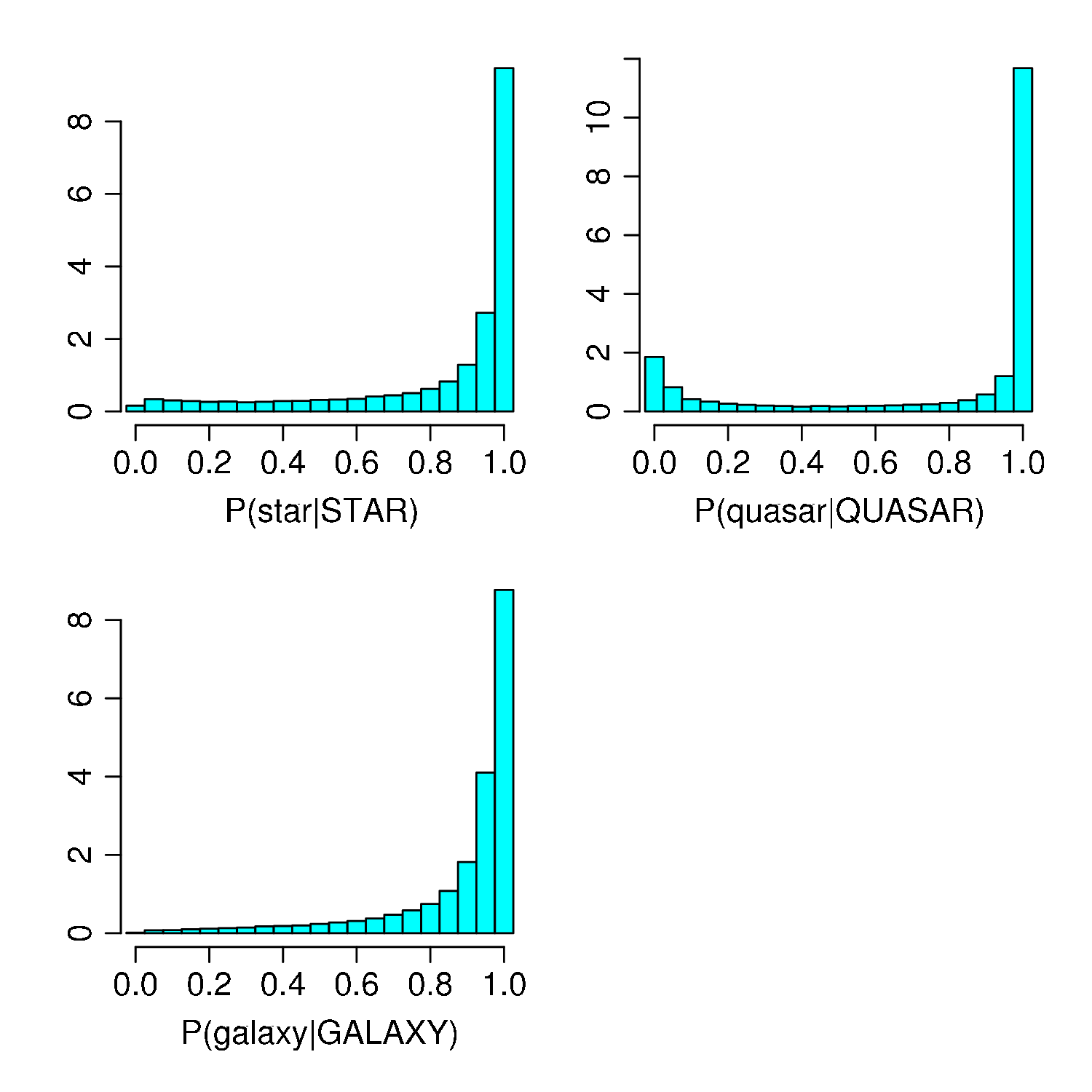}
\caption[Histograms $P({\tt class} | {\tt CLASS})$, nominal priors,
G=18.5]{Histograms of DSC outputs for each class showing how
  confident DSC is of identifying each class. Nominal
  priors, G=20.0, no low EW quasars in training data
\label{fig:nom200_correct_class_histogram}}
\end{center}
\end{figure}

\begin{figure} 
\begin{center}
\includegraphics[width=0.49\textwidth]{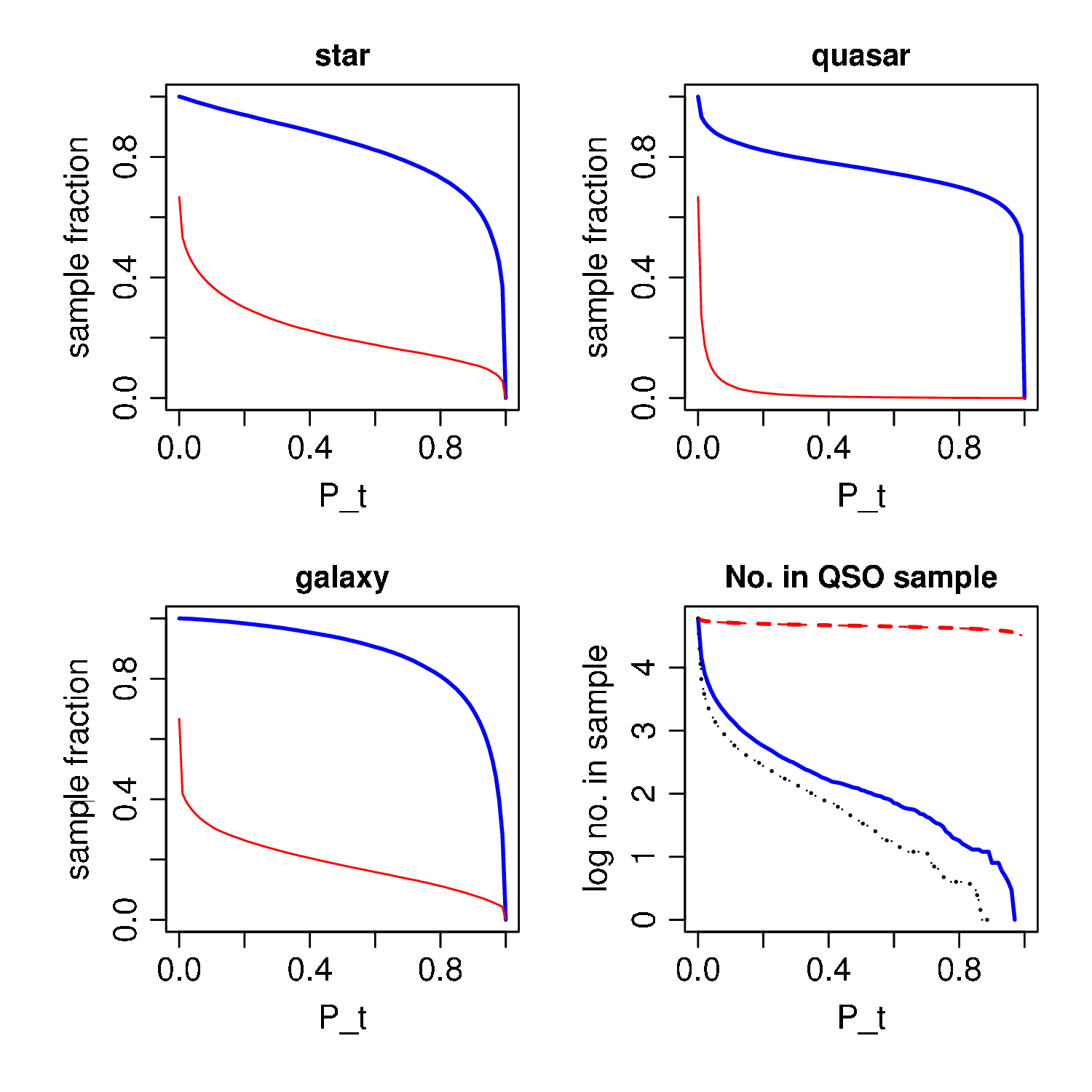}
\caption[Completeness \& contamination, nominal priors,
G=20.0]{Completeness (blue/thick line) and contamination (red/thin line)
of a sample as a function of the probability threshold. 
The bottom right panel shows the (logarithm) of the actual number (thick lines) of different types of class in the quasar sample (i.e.\ in the test set): stars (blue/solid line); quasars (red/dashed lines).
Nominal priors, G=20.0, no low EW quasars in training data
\label{fig:nom200_completeness_confusion}}
\end{center}
\end{figure}

\begin{figure} 
\begin{center}
\includegraphics[width=0.49\textwidth]{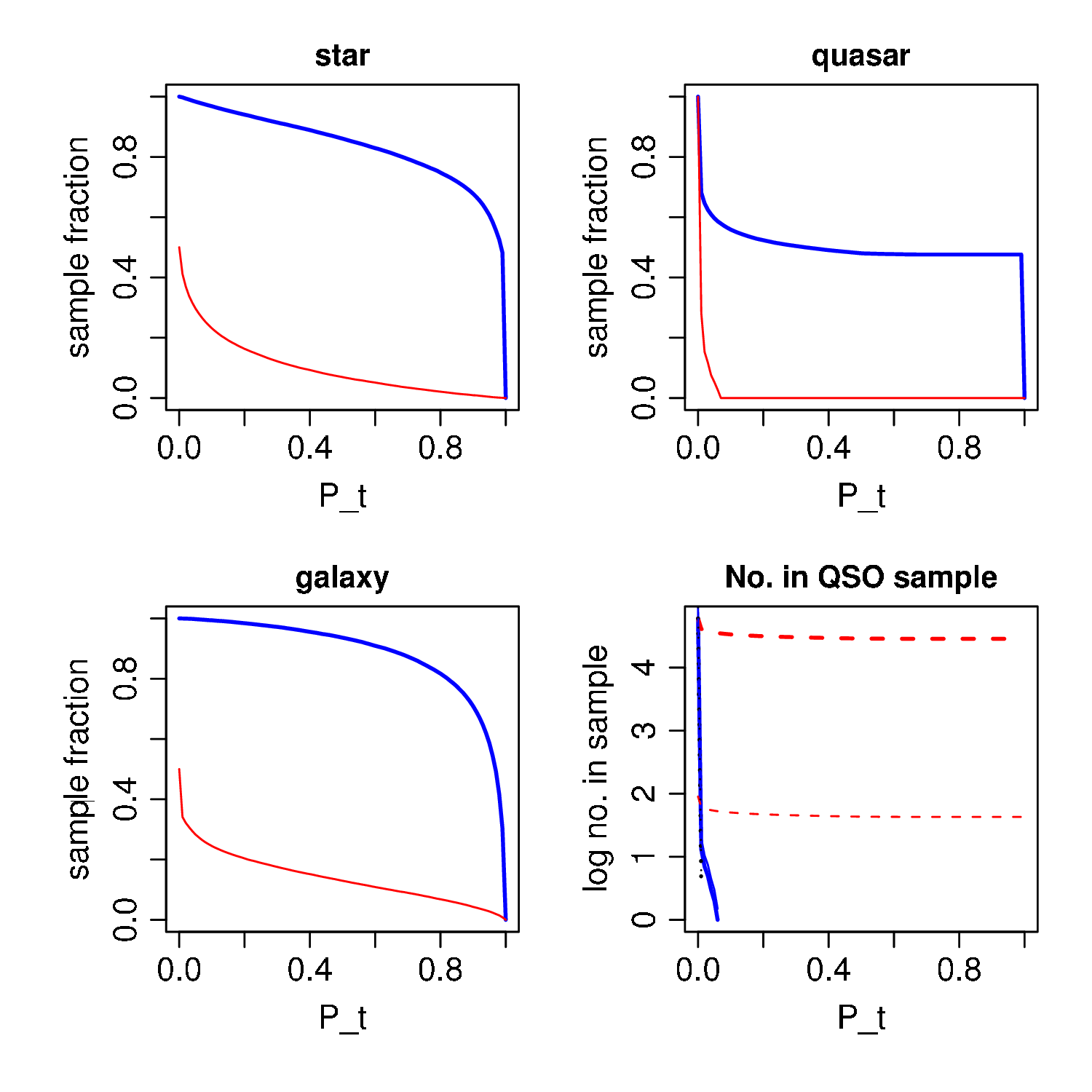}
\caption[Completeness \& contamination, modified priors,
G=20.0]{Completeness (blue/thick line) and contamination (red/thin
  line) of a sample as a function of the probability threshold.  The
  bottom right panel shows the (logarithm) of the actual number (thick
  lines) of different types of class in the quasar sample (i.e.\ in
  the test set): stars (blue/solid line); quasars (red/dashed
  lines). The thin lines show the corresponding {\em effective} number
  of objects for the modified case. There are no galaxy contaminants.
  Modified priors, G=20.0, no low EW quasars in training data
\label{fig:mod200_completeness_confusion}}
\end{center}
\end{figure}

At G=20 the noise is higher, so we expect DSC to perform less
well. The basic class assignment based on maximum posterior class
probability is indeed worse
(Table~\ref{tab:nom200_contingency_table}). Comparing with
Table~\ref{tab:nom185_contingency_table} we see that the stars suffer
in particular, with only 86\% being correctly classified, as opposed to
99\% at G=18.5. This is reflected by the lower confidence the model
gives to its classifications
(Fig.~\ref{fig:nom200_correct_class_histogram}).  Likewise, the
completeness is lower and the contamination higher at a given threshold
(Fig.~\ref{fig:nom200_completeness_confusion}).

If we now modify the priors and recalculate the posterior
probabilities then we see a similar shift in the distributions that we
did for the G=18.5 case (histograms not shown). The C\&C plot with the
modified model is shown in
Fig.~\ref{fig:mod200_completeness_confusion}. What stands out
immediately is that we are still able to get a clean (zero
contamination) quasar sample. This occurs once $P_t = 0.07$, at which
point the completeness is 58\% (and remains at above 50\% if we choose
to take a larger threshold). This compares to a completeness of 65\%
at $P_t=0.11$ for zero contamination at $G=18.5$. Thus we see that the
higher noise in the data can be translated into a lower completeness
(which is acceptable) rather than a raised contamination (which is
not). Notice that there is no guarantee that the contamination should
drop to zero before $P_t=1$ (cf.\ section~\ref{g185}), so this is an
important result.

\subsection{The impact of using astrometry in classification}

We may expect parallaxes and proper motions to provide a good
discriminant between Galactic and extragalactic objects. Yet it turns
out that if we remove astrometry from the DSC, the performance hardly
degrades (accuracy decreases by no more than 1\%).  This observation
may not be surprising, because the majority of objects with astrometry
consistent with zero (within the measurement errors) are actually
stars, and that is because our training sample is dominated by distant
stars. It is also not specific to the SVM. We also built a two-stage
classifier, in which (1) an SVM predicts classes based only on
photometry and (2) Gaussian mixtures are used to model the density in
the 2D astrometric space and predict classes. The probabilities are
then combined to give a single posterior (Bailer-Jones \&
Smith~\cite{cbj08b}). The results are no better than an SVM using
BP/RP and astrometry (although the two-stage classifier offers more
flexibility and interpretability).

\subsection{The effect of reducing the fraction of quasars in the training data}\label{reduceqsos}

In section~\ref{traininf} we argued that the model class priors are
not necessarily dictated by the class fractions in the training data.
Moreover, in attempting to match the training data distribution to
that in the target population for a rare object, we may end up with
very few objects in the training data, such that the classifier does
not learn to classify them well.

\begin{figure} 
\begin{center}
\includegraphics[width=0.30\textwidth]{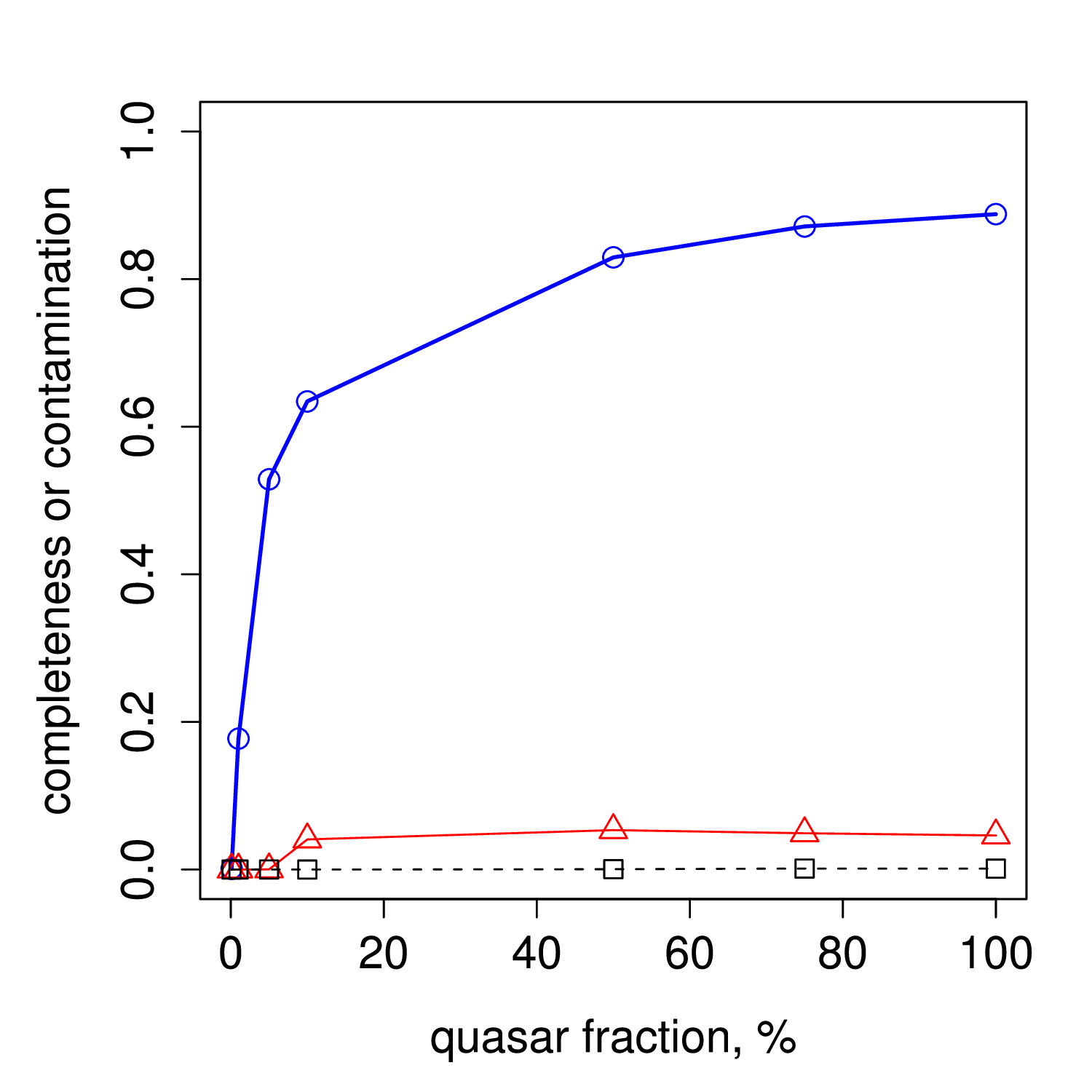}
\caption[]{Quasar sample completeness (thick blue line) and
  contamination from stars (thin red line) and galaxies (dotted black
  line) as a function of varying the fraction of quasars in the
  training data set (the first two values are 0.1\% and 1\%)
  \label{removeqsos}}
\end{center}
\end{figure}

We test this by varying the fraction of quasars in the training data
set, $q$, from 100\% (equal class fractions) down to 0.1\% (60
quasars). We then build samples based on maximum probability and
measure the C\&C (again using equation~\ref{eqn:mod_compcont} to
account for the modified class fractions).  Fig.~\ref{removeqsos}
shows how the completeness and contamination of the quasar sample
varies with $q$.  For modest reductions in $q$ the contamination is
5\%, which is too high for a pure sample (although the samples are
built using maximum probability, so we could reduce this at the cost
of reduced completeness).  Surprisingly, the contamination actually
drops to zero once the quasar fraction drops below 5\%. However, this
comes at the price of greatly reduced completeness of just 0.1\% at
$q=$\,0.1\%. This we can directly compare to the results from the
modified model in experiment (2) (section~\ref{modmodel}),
where the completeness is still 62\%.  Thus we conclude that pruning
the training data set is not the way to achieve a good classifier for
a rare class.

\section{Discussion}\label{discussion}

\subsection{Training data distributions and class definitions}

Any supervised method is limited by the data it is trained on.
Systematic differences between these and the target population can
compromise the classifier.  It is not the goal of this article to
present the final classifier for Gaia, so we have not yet optmized the
training sets. Nonetheless, our work does raise some issues relevant
to this important next step, which we briefly discuss.

First, we found that removing the low EW quasars was important for
achieving clean quasar samples in the modified case.  This may reflect
a limitation of the SVMs, but it raises the general issue of how to
set the training data distribution. We do not expect a supervised
method to perform well on regions of the data space not included in
the training data (although we did find that the model could correctly
classify many low EW quasars). For this reason, DSC will be preceeded
by an ``outlier detector'' in the classification pipeline. This assesses
how well an observed object is ``covered'' by the training space and
only passes ``known'' objects to DSC. It could be based on density
estimation (in low dimensions) or one-class SVMs.

Second, if we are building a classifier only to look for a specific
type of object, then we don't need to aim for completeness in the
training sample (although we do need to include potential
contaminants).  For example, none of our quasar simulations include
interstellar extinction (unlike the stars and galaxies).  This could
be motivated by saying we do not trust extinction laws and just want
to find unreddened quasars. We are free to make this choice at the
cost of reduced completeness of real quasars.  If we included reddened
quasars in the test set, then either they would be classified as
quasars anyway (in which case the completeness goes up), or they would
be classified as something else. Either way, the quasar contamination
would not increase so our conclusion about building pure samples of
unreddened quasars is unaffected. Incidentally, the results in
section~\ref{results} are very similar if we repeat the experiments
but now with interstellar extinction applied also to the quasars. The
main difference is that the quasar sample completeness is reduced by
around 10\%, with the more reddened quasars now contaminating the star
samples.  The quasar contamination is not increased.

Third, how should we define the classes? If we had split the quasars
into two classes (e.g.\ high and low emission line EW) maybe
this would help the SVM to retain zero contamination on the high EW
class without having to remove the low EW quasars entirely from the
training data. This will be the subject of future work.

Gaia does not exist in a void; we of course have information from
other catalogues/surveys to help us identify quasars.  The full DSC is
actually a multi-stage algorithm, in which the present paper just
describes the stage based on the BP/RP spectrum. Other stages provide
class probabilities based on other data, for example the source
magnitude or Galactic latitute or external catalogues, in which case
we might call them ``priors''. We will describe this multi-stage
approach in a future paper.

\subsection{How do we interpret posterior probabilities?}\label{interp}

When we build a sample by applying a threshold, all of the objects
have a posterior probability which is at least as high as the
threshold. For example, with the modified model in
section~\ref{modmodel}, we were able to set $P_t=0.11$ and produce a
sample of only quasars. There is of course no guarantee that all
samples built using this threshold have no contaminants.  On the other
hand, this threshold is not a statement about the sample as a whole:
it would be wrong to expect only 11\% of the sample to be quasars.  We
need to look at the actual proabilities, not a lower limit.  How can
we use these?  Imagine a case in which we set $P_t=0.90$ and
thereby obtained a sample of 1000 (supposed) quasars, all which happen
to have $P({\tt quasar})=0.95$. We would reasonably expect 5\% not
to be quasars.  Yet we must be careful with such an inference, because
posterior probabilities are not frequencies (especially if we have
very non-uniform priors, e.g.\ class fractions of $(1, 0.001, 1)$).

\begin{figure}
\begin{center}
\includegraphics[width=0.50\textwidth]{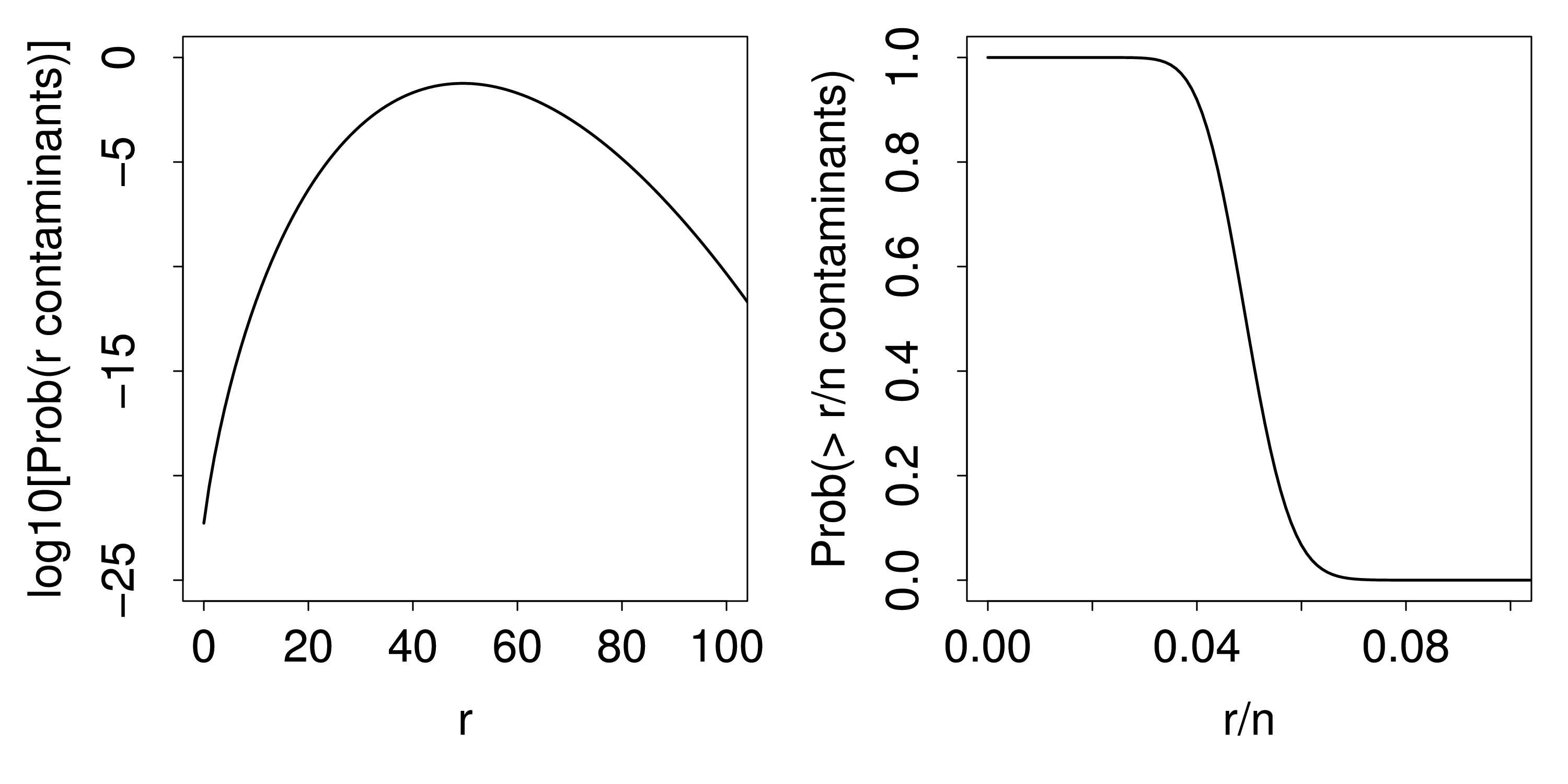}
\caption[Sample contamination probability]{The left panel shows the
  variation with $r$ of $P(n=1000,r,p=0.95)$
  (equation~\ref{sampleprob}), the probability that there are exactly
  $r$ contaminants in the quasar sample of $n$ objects, each with
  posterior quasar probability of $p$. The right hand panel is 1 minus
  the cumulation of this (equation~\ref{sampcont}), and so shows the
  probability that there will be more than $r$ contaminants (shown as
  the contamination fraction $r/n$).
\label{fig:samplepr}}
\end{center}
\end{figure}

How can we make statements about the sample as a whole from the
individual probabilities?  The general case is complex, but for the
case that all the probabilities are the same, the solution is simple.
If $P({\tt quasar}) =
p$ for all objects in a sample of size $n$, then the probability that
exactly $r$ of them ($r \leq n$) are non-quasars is
\begin{equation}
P(n,r,p) = \frac{n!}{r!(n-r)!}(1-p)^r p^{(n-r)}
\label{sampleprob}
\end{equation}
The left panel of Fig.~\ref{fig:samplepr} shows how (the logarithm of)
this quantity varies with $r$ for $n=1000$ and $p=0.95$.  (It has a
maximum of $P(n,r,p)=0.058$ at $r=50$). By summing
equation~\ref{sampleprob} over ranges of $r$ we can make more useful
statements. For example, summing from $r=0$ to $r=50$, we can say that
the probability that there are 50/1000=5\% contaminants or fewer is
0.54. Equivalently, the probability that there is more than 5\%
contamination is 0.46.  The right hand panel of
Fig.~\ref{fig:samplepr} shows this for variable $r$. Specifically it plots
\begin{equation}
1 - \sum_{r'=0}^{r'=r} P(n,r',p)
\label{sampcont}
\end{equation}
against $r/n$, the contamination fraction. This figure agrees approximately 
with our intuition of a 5\% contamination when all posterior
probabilities are 95\%.

We can apply this to the case in section~\ref{testpred}, where we
applied the modified model to a test with 60 quasars. Setting
$P_t=0.2$ we achieved a zero contamination sample of 37 quasars (62\%
completeness; Table~\ref{tab:mod185_contingency_table}). If we didn't
know the true classes, would we have expected zero contamination with
the above calculation?  The posterior probabilities are not equal (the
expected distribution is $P({\tt quasar} | {\tt QUASAR})$ in
Fig.~\ref{fig:nom185_all_histogram}), but we approximate by setting
$p$ equal to the average for this sample, which is
0.95. Equation~\ref{sampcont} tells us ($n=37$, $r=1$, $p=0.95$) that
the probability of having one or more contaminants is 0.56. This is
consistent with having no contaminants. Of course, this average
probability is not very representative, so this doesn't strictly
apply. For example, we could also have taken a threshold of $P_t=0.50$
and still had a zero contamination sample, but in this case the
average is $p=0.997$ and the probability of one or more contaminants
is just 0.006.

Despite these approximations, it appears that the inferences about the
sample as a whole are consistent with the individual posterior
probabilities.

\subsection{Why we should prefer the modified model}

Ultimately we would like to know whether the modified model is
``better'' than the nominal one. If we ignored the issue of
priors and class fractions, then our predictions of sample
completeness and contamination would always be given by
Fig.~\ref{fig:nom185_completeness_confusion} (at G=18.5), regardless of the true
class fractions in the target population.  If we
improve this by instead using equation~\ref{eqn:mod_compcont} to
accommodate the modified class fractions (but still with the nominal
model posteriors), then our predictions would be the solid lines in
Fig.~\ref{fig:nom_cc_comparison}. Yet both are poor predictions of the
actual completeness and contamination in a sample where quasars are
rare (dashed lines in Fig.~\ref{fig:mod_cc_comparison}).

We demonstrated (in section~\ref{nom_limited}) that the application of
the nominal model to a data set with rare quasars gives poorer results
(larger contamination) than the modified model {\em at a given
  probability threshold}. It is not yet clear whether we could just
use the nominal model but with much higher probability thresholds to
achieve similar results. However, there are two reasons why we should
not want to do this.  First, as we must not build test data sets with
very small numbers of the rare class, we would still have to use the
modified class fractions in order to correctly predict the C\&C
(section~\ref{modcalcs}). Second, we want models which deliver actual
probabilities, not just numbers between 0.0 and 1.0 to which we apply
a meaningless threshold. This is especially important if we later want
to update the probabilities, e.g.\ in a multi-stage classifier. We
therefore believe it is better to use the modified model.


\section{Conclusions}\label{conclusions}

We have introduced a method of probability modification which can be used to
build clean samples of rare objects for which the completeness and
contamination can be reliably predicted.  We have demonstrated this
using a support vector machine classifier, although it may, of course,
be used in conjunction with any classifier which gives probabilities.
The main conclusions of our work are as follows.

\begin{itemize}

\item To construct a pure sample of objects we should use a
  probabilistic classifier and only select objects with high
  probabilities. By varying this probability threshold we can
  trade off sample completeness and contamination.

\item To achieve pure samples of rare objects, we must take into
  account the expected class fractions in the target population, which
  act as a prior probability on the classifier. We use these to modify
  the nominal classifier outputs to give the {\em modified model}.

\item We applied our modified model to a three class problem in which
  quasars are simulated to be 1000 times rarer than stars and
  galaxies.  We can achieve a pure quasar sample (zero contamination)
  yet still reach a sample completeness of 50--65\% for 
  magnitudes down to G=20.0. Although the test set is finite in size,
  this correponds to an upper limit in the contamination of 1 in
  39\,000.  This is more than adequate for establishing an astrometric
  reference frame for Gaia: If Gaia observes half a million quasars,
  we can build a quasar sample of 250\,000 with no more than 13
  contaminants.

\item While achieving this pure quasar sample, we simultanesouly
  achieve very complete galaxy and star samples (both 99\%) with low
  contamination (both 0.7\%) (figures for G=18.5). 

\item These results were achieved after removing quasars with low
  equivalent width emission lines from the training sample (defined
  somewhat arbitrarily as 5000\,\AA). Including these precluded
  establishing a low contamination sample, because it resulted in cool
  (4000--8000\,K) highly reddened (\av\,=\,8--10) stars being confused
  with them. After removing these quasars, the first stars to
  contaminate a quasar sample (if we set a low threshold) are these
  reddened cool stars as well as hot stars (\teff\,$>$\,40\,000\,K).

\item Including parallax and proper motion (either as
  additional SVM inputs, or in a separate mixture model classifier)
  hardly changes the performance.  This is not surprising since the
  majority of objects with astrometry consistent with zero are
  actually stars.

\item Extending the training and testing sets to include quasars with
  a full range of interstellar extinction does not significantly alter
  the results (completeness slightly lower, but contamination
  unaffected)

\item All classification models have a prior, but the prior is often
  not explicit and are sometimes implicitly influenced by the training
  data distribution. We have introduced a simple method for
  calculating the implicit priors in a classification model, which we
  call the {\em model-based priors}.  In many cases we have
  experimented with, these priors are close to the true class
  fractions in the training data (nominal model) or modified class
  fractions (modified model).

\item We recommend that a classifier be trained on roughly equal
  numbers of objects in each class so that it can properly learn the
  class distributions or boundaries. By determining the model-based
  priors and replacing them with something more appropriate to the
  target population (e.g.\ quasars being rare), we can produce a
  modified model with superior performance. In particular, this is far
  better at producing large, pure samples of the rare
  class. 

\item With our approach we can apply the model to any new target
  population by specifying the appropriate class fractions (priors)
  without having to change the training data distribution or re-train
  the model.

\end{itemize}

\section*{Acknowledgements}

This work makes use of Gaia simulated observations and we thank the
members of the Gaia DPAC Coordination Unit 2, in particular Paola
Sartoretti and Yago Isasi, for their work.  These data simulations were
done with the MareNostrum supercomputer at the Barcelona
Supercomputing Center - Centro Nacional de Supercomputaci\'on (The
Spanish National Supercomputing Center).  We thank Jean-Francois
Claeskens, Vivi Tsalmantza and Jean-Claude Bouret for use of the
quasar, galaxy and OB star spectral libraries (respectively), and
Christian Elting for assistance in assembling the data samples.  The
MPIA Gaia team was supported in part by a grant from the Deutsches
Zentrum f\"ur Luft- und Raumfahft (DLR).


\end{document}